# Introduction to Chiral Symmetry

Volker Koch

*Nuclear Science Division, Lawrence Berkeley National Laboratory*
*Berkeley, CA, 94720, U.S.A.*

## Abstract

These lectures are an attempt to a pedagogical introduction into the elementary concepts of chiral symmetry in nuclear physics. Effective chiral models such as the linear and nonlinear sigma model will be discussed as well as the essential ideas of chiral perturbation theory. Some applications to the physics of ultrarelativistic heavy ion collisions will be presented.

# 1  Introduction

Chiral symmetry is a symmetry of QCD in the limit of vanishing quark masses. We know, however, that the current quark masses are finite. But compared with hadronic scales the masses of the two lightest quarks, up and down, are very small, so that chiral symmetry may be considered an approximate symmetry of the strong interactions.

Long before QCD was known to be the theory of strong interactions, phenomenological indications for the existence of chiral symmetry came from the study of the nuclear beta decay. There one finds, that the weak coupling constants for the vector and axial-vector hadronic-currents, $C_V$ and $C_A$, did not (in case of $C_V$) or only slightly (25% in case of $C_A$) differ from those for the leptonic counterparts. Consequently strong interaction 'radiative' corrections to the weak vector and axial vector 'charge' are absent. The same is true for the more familiar case of the electric charge, and there we know that it is its conservation, which protects it from radiative corrections. Analogously, we expect the weak vector and axial vector charge, or more generally, currents, to be conserved due to some symmetry of the strong interaction. In case of the vector current, the underlying symmetry is the well known isospin symmetry of the strong interactions and thus the hadronic vector current is identified with the isospin current. The identification of the axial current, on the other hand is not so straightforward. This is due to another, very important and interesting feature of the strong interaction, namely that the symmetry associated with the conserved axial vector current is 'spontaneously broken'. By that, one means that while the Hamiltonian possesses the symmetry, its ground state does not. An important consequence of the spontaneous breakdown of a symmetry is the existence of a massless mode, the so called Goldstone-boson. In our case, the Goldstone boson is the pion. If chiral symmetry were a perfect symmetry of QCD, the pion should be massless. Since chiral symmetry is only approximate, we expect the pion to have a finite but small (compared to all other hadrons) mass. This is indeed the case!

The fact that the pion is a Goldstone boson is of great practical importance. Low energy/temperature hadronic processes are dominated by pions and thus all observables can be expressed as an expansion in pion masses and momenta. This is the basic idea of chiral perturbation theory, which is very successful in describing threshold pion physics.

At high temperatures and/or densities one expects to 'restore' chiral symmetry. By that one means, that, unlike the ground state, the state at high temperature/density posses the same symmetry as the Hamiltonian (the symmetry of the Hamiltonian of course will not be changed). As a consequence of this so called 'chiral restoration' we expect the absence of any Goldstone modes and thus the pions, if still present, should become as massive as all other hadrons[1]. To create a system of restored chiral symmetry

---
[1] If of course chiral restoration and deconfinement take place at the same temperature, as current lattice gauge calculations suggest, the concept of hadrons in the restored phase may become meaningless.

in the laboratory is one of the major goals of the ultra-relativistic heavy ion experiments.

These lectures are intended to serve as an introduction into the ideas of chiral symmetry in particular for experimentalists interested or working in this field. Thus emphasis will be put on the ideas and concepts rather than formalism. Consequently, most arguments presented will be heuristic and/or based on simple effective models. References will be provided for those seeking more rigorous derivations.

In the first section we will introduce some basic concepts of quantum field theory, which are necessary to discuss the effect of symmetries on the dynamics. Then we will introduce the chiral symmetry transformations and derive some results, such as the Goldberger-Treiman relation. In the second section we will present the linear sigma model as the most simple effective chiral model. Using this rather intuitive model we will discuss explicit chiral symmetry breaking. As an application we will consider pion-nucleon scattering. The third section will be devoted to the so called nonlinear sigma model, which then serves as a basis for the introduction into chiral perturbation theory. In the last section we will give some examples for chiral symmetry in the physics of hot and dense matter.

Preparing these lectures I have borrowed from many sources. Those which I personally found most useful are listed at the end of this contribution. This is certainly a personal choice as there are many other books and articles on subject available. If not stated otherwise, the conventions of Bjorken and Drell [1] are used for metric, gamma-matrices etc.

## 2 Theory Primer

### 2.1 Basics of quantum field theory

Quantum field theory is usually written down in the Lagrangian formulation (see e.g. the book of Bjorken and Drell [1]). Let's start out with what we know from classical mechanics. There, one obtains the equations of motion from the Hamilton principle, where one requires that the variation of the action $S = \int dt\, L(q, \dot{q}, t)$ vanishes

$$\delta S = 0 \;\;\Rightarrow\;\; \frac{d}{dt}\frac{\partial L}{\partial \dot{q}} - \frac{\partial L}{\partial q} = 0 \tag{1}$$

Here $S$ is called the action and $L = T - V$ is the Lagrange-function. For example, Newtons equations of motion for a particle in a potential $V(q)$ derive from

$$L = \frac{1}{2}m\dot{q}^2 - V(q) \tag{2}$$

$$\Rightarrow m\ddot{q} + \frac{\partial V}{\partial q} = 0 \;\;\Leftrightarrow\;\; m\ddot{q} = -\frac{\partial V}{\partial q} = F \tag{3}$$

If one goes over to a field theory, the coordinates $q$ are replaced by the fields $\Phi(x)$ and the velocities $\dot{q}$ are replaced by the derivatives of the fields

$$q \;\rightarrow\; \phi(x) \tag{4}$$

$$\dot{q} \;\rightarrow\; \partial_\mu \Phi(x) \equiv \frac{\partial \Phi(x)}{\partial x^\mu} \tag{5}$$

and the Lagrange-function is given by the spatial integral over the Lagrangian density, $\mathcal{L}$, or Lagrangian, as we shall call it from now on

$$L = \int d^3x\, \mathcal{L}(\Phi(x), \partial_\mu \Phi(x), t) \tag{6}$$

$$S = \int dt\, L = \int d^4x\, \mathcal{L}(\Phi(x), \partial_\mu \Phi(x), t) \tag{7}$$

Lorentz invariance implies that the action $S$ and thus the Lagrangian $\mathcal{L}$ transform like Lorentz-scalars. The equations of motion for the fields are again obtained by requiring that the variation of the action S vanishes. This variation is carried out by a variation of the fields

$$\Phi \;\rightarrow\; \Phi + \delta\Phi \tag{8}$$

$$\partial_\mu \Phi \;\rightarrow\; \partial_\mu \Phi + \delta(\partial_\mu \Phi) \tag{9}$$

with

$$\delta(\partial_\mu \Phi) = \partial_\mu(\Phi + \delta\Phi) - \partial_\mu \Phi = \partial_\mu(\delta\Phi) \qquad (10)$$

Consequently,

$$\begin{aligned} \delta S &= \int d^4x\, \mathcal{L}(\Phi + \delta\Phi, \partial_\mu\Phi + \delta(\partial_\mu\Phi)) - \mathcal{L}(\Phi, \partial_\mu\Phi) \\ &= \int d^4x\, \mathcal{L}(\Phi, \partial_\mu\Phi) + \frac{\partial \mathcal{L}}{\partial \Phi}\delta\Phi + \frac{\partial \mathcal{L}}{\partial(\partial_\mu\Phi)}\delta(\partial_\mu\Phi) - \mathcal{L}(\Phi, \partial_\mu\Phi) \\ &= \int d^4x\, \frac{\partial \mathcal{L}}{\partial \Phi}\delta\Phi + \frac{\partial \mathcal{L}}{\partial(\partial_\mu\Phi)}\partial_\mu(\delta\Phi) \end{aligned} \qquad (11)$$

where equ. (10) has been used. The derivatives of $\mathcal{L}$ with respect to the fields are so called functional derivatives, but for all practical purposes they just work like 'normal' derivatives, where $\mathcal{L}$ is considered a function of the fields $\Phi$. Partial integration of the second term of equ. (11) finally gives

$$0 = \delta S = \int d^4x \left( \frac{\partial \mathcal{L}}{\partial \Phi} - \partial_\mu(\frac{\partial \mathcal{L}}{\partial(\partial_\mu\Phi)}) \right) (\delta\Phi) \qquad (12)$$

which leads to the following equations of motion, since the variation $\delta\Phi$ are arbitrary

$$\frac{\partial \mathcal{L}}{\partial \Phi} - \partial_\mu(\frac{\partial \mathcal{L}}{\partial(\partial_\mu\Phi)}) = 0 \qquad (13)$$

If we are dealing with more than one field, such as in case of pions, where we have three different charge states, the equations of motion have the same form as in equ. (13) only that the fields carry now an additional index labeling the different fields under consideration

$$\frac{\partial \mathcal{L}}{\partial \Phi_i} - \partial_\mu(\frac{\partial \mathcal{L}}{\partial(\partial_\mu\Phi_i)}) = 0 \qquad (14)$$

As example let us consider the Lagrangian of a free boson and fermion field respectively.

**(i)** free scalar bosons of mass $m$:

$$\mathcal{L}_{K.G.} = \frac{1}{2}(\partial_\mu\Phi\partial^\mu\Phi) - \frac{1}{2}m^2\Phi^2 \qquad (15)$$

$$\Rightarrow \frac{\partial \mathcal{L}}{\partial \Phi} = -m^2\Phi \qquad (16)$$

$$\Rightarrow \partial_\mu(\frac{\partial \mathcal{L}}{\partial(\partial_\mu\Phi_i)}) = \partial_\mu\partial^\mu\Phi \qquad (17)$$

Thus, according to equ. (14) the equation of motion is

$$(\partial_\mu \partial^\mu + m^2)\Phi = (\partial_t^2 - \nabla^2 + m^2)\Phi = 0 \tag{18}$$

which is just the well known Klein-Gordon equation for a free boson.

**(ii)** free fermions of mass $m$

$$\mathcal{L}_{\mathcal{F.D.}} = \bar{\psi}(i\gamma_\mu \partial^\mu - m)\psi \tag{19}$$

By using the conjugate field $\bar{\psi}$ in the equation of motion (14)

$$\Rightarrow \frac{\partial \mathcal{L}}{\partial \bar{\psi}} = (i\gamma_\mu \partial^\mu - m)\psi \tag{20}$$

$$\Rightarrow \frac{\partial \mathcal{L}}{\partial(\partial_\mu \bar{\psi})} = 0 \tag{21}$$

we obtain the Dirac equation for $\psi$ :

$$(i\gamma^\mu \partial_\mu - m)\psi = 0 \tag{22}$$

whereas inserting $\psi$ for $\Phi_i$ in (14) leads to the conjugate Dirac equation

$$\bar{\psi}(i\gamma^\mu \overleftarrow{\partial}_\mu + m) = 0 \tag{23}$$

## 2.2 Symmetries

One of the big advantages of the Lagrangian formulation is that symmetries of the Lagrangian lead to conserved quantities (currents). In classical mechanics we know that symmetries of the Lagrange function imply conserved quantities. For example, if the Lagrange function is independent of space and time, momentum and energy are conserved, respectively.

Let us assume that L is symmetric under a transformation of the fields

$$\Phi \longrightarrow \Phi + \delta\Phi \tag{24}$$

meaning

$$\mathcal{L}(\Phi + \delta\Phi) = \mathcal{L}(\Phi) \tag{25}$$

$$\Rightarrow 0 = \mathcal{L}(\Phi + \delta\Phi) - \mathcal{L}(\Phi) = \frac{\partial \mathcal{L}}{\partial \Phi}\delta\Phi + \frac{\partial \mathcal{L}}{\partial(\partial_\mu \Phi)}\delta(\partial_\mu \Phi) \tag{26}$$

where we have expanded the first term to leading order in $\delta\Phi$. Using equ. (10) and the equation of motion (13) we have

$$\begin{aligned} 0 &= \left(\partial_\mu \frac{\partial \mathcal{L}}{\partial \Phi}\right) \delta\Phi + \frac{\partial \mathcal{L}}{\partial(\partial_\mu \Phi)} (\partial_\mu \delta\Phi) \\ &= \partial_\mu \left(\frac{\partial \mathcal{L}}{\partial(\partial_\mu \Phi)} \delta\Phi\right) \end{aligned} \qquad (27)$$

so that

$$J_\mu = \frac{\partial \mathcal{L}}{\partial(\partial_\mu \Phi_i)} \delta\Phi_i \qquad (28)$$

is a conserved current, with $\partial^\mu J_\mu = 0$. In the last equation we have included the indices for possible different fields $\Phi_i$.

As an example, let us discuss the case of a unitary transformation on the fields, such as e.g. an isospin rotation among pions. For obvious reasons unitary transformations are the most common ones, and the chiral symmetry transformations also belong to this class.

$$\Phi_i \longrightarrow \Phi_i - i\Theta_a T^a_{ij} \Phi_j \qquad (29)$$

where $\Theta$ corresponds to the rotation angle and $T^a_{ij}$ is a matrix, usually called the generator of the transformation (isospin matrix in case of isospin rotations). The index $a$ indicates that there might be several generators associated with the symmetry transformation (in case of isospin rotations, we have three isospin matrices). Equation (29) corresponds to the expansion for small angles of the general transformation

$$\vec{\Phi} \longrightarrow e^{-i\Theta^a \hat{T}^a} \vec{\Phi} \qquad (30)$$

where the vector on $\vec{\Phi}$ indicates the several components of the field $\Phi$ such as $\pi^+$, $\pi^-$ and $\pi^0$. From equ. (28) and equ. (29) we find the following expression for the conserved currents

$$J^a_\mu = -i \frac{\partial \mathcal{L}}{\partial(\partial_\mu \Phi_j)} T^a_{jk} \Phi_k \qquad (31)$$

where we have divided by the angle $\Theta^a$. This current is often referred to as a Noether current, after E. Noether who first showed its existence[2].

---

[2] Note, that some of the Noether currents are not conserved on the quantum-level. In other word, not every symmetry of the classical field theory has a quantum analog. If this is not the case one speaks of anomalies. For a discussion of anomalies, see e.g. [15].

Of course, a conserved current leads to a conserved charge

$$Q = \int d^3x J_0(x); \quad \frac{d}{dt}Q = 0 \tag{32}$$

Finally, let us add a small symmetry breaking term to the Lagrangian

$$\mathcal{L} = \mathcal{L}_0 + \mathcal{L}_1 \tag{33}$$

where $\mathcal{L}_0$ is symmetric with respect to a given symmetry transformation of the fields and $\mathcal{L}_1$ breaks this symmetry. Consequently, the variation of the Lagrangian $\mathcal{L}$ does not vanish as before but is given by

$$\delta\mathcal{L} = \delta\mathcal{L}_1 \tag{34}$$

Following the steps above, we can easily convince ourselves, that the variation of the Lagrangian can still be expressed as the divergence of a current, which is given by equ. (28) or (31), in case of unitary transformations of the fields. Thus we have

$$\delta\mathcal{L} = \delta\mathcal{L}_1 = \partial^\mu J_\mu \tag{35}$$

Since $\delta\mathcal{L}_1 \neq 0$ the current $J_\mu$ is not conserved. Relation (35) nicely shows how the symmetry breaking term of the Lagrangian is related to the non-conservation of the current. It will also prove very useful when we later on introduce the slight breaking of chiral symmetry due to the finite quark masses.

### 2.2.1 Example: Massless fermions

As an example for the Noether current, let us consider the Lagrangian of two flavors of massless fermions. Since we will only discuss transformations on the fermions, the results will be directly applicable to massless QCD.

The Lagrangian is given by (see eq. (19))

$$\mathcal{L} = i\bar{\psi}_j \partial\!\!\!/\psi_j \tag{36}$$

where the index '$j$' refers to the two different flavors, let's say 'up' and 'down', and $\partial\!\!\!/$ is the usual shorthand for $\partial_\mu \gamma^\mu$.

**(i)** Consider the following transformation

$$\Lambda_V : \psi \longrightarrow e^{-i\frac{\vec{\tau}}{2}\vec{\Theta}}\psi \simeq (1 - i\frac{\vec{\tau}}{2}\vec{\Theta})\psi \tag{37}$$

where $\vec{\tau}$ refers to the Pauli - (Iso)spin- matrices, and where we have switched to a iso-spinor notation for the fermions, $\psi = (u, d)$. The conjugate field, $\bar{\psi}$ transforms under $\Lambda_V$ as follows

$$\bar{\psi} \longrightarrow e^{+i\frac{\vec{\tau}}{2}\vec{\Theta}} \bar{\psi} \simeq (1 + i\frac{\vec{\tau}}{2}\vec{\Theta})\bar{\psi} \tag{38}$$

and, hence, the Lagrangian is invariant under $\Lambda_V$

$$i\bar{\psi}\slashed{\partial}\psi \longrightarrow i\bar{\psi}\slashed{\partial}\psi - i\vec{\Theta}\left(\bar{\psi}i\slashed{\partial}\frac{\vec{\tau}}{2}\psi - \bar{\psi}\frac{\vec{\tau}}{2}i\slashed{\partial}\psi\right)$$

$$= i\bar{\psi}\slashed{\partial}\psi \tag{39}$$

Following equ. (31) the associated conserved current is

$$V_\mu^a = \bar{\psi}\,\gamma_\mu\frac{\tau^a}{2}\,\psi \tag{40}$$

and is often referred to as the 'vector-current'.

**(ii)** Next consider the transformation

$$\Lambda_A: \quad \psi \longrightarrow e^{-i\gamma_5\frac{\vec{\tau}}{2}\vec{\Theta}}\psi = (1 - i\gamma_5\frac{\vec{\tau}}{2}\vec{\Theta})\psi \tag{41}$$

$$\Rightarrow \ \bar{\psi} \longrightarrow e^{-i\gamma_5\frac{\vec{\tau}}{2}\vec{\Theta}}\bar{\psi} \simeq (1 - i\gamma_5\frac{\vec{\tau}}{2}\vec{\Theta})\bar{\psi} \tag{42}$$

where we have made use of the anti-commutation relations of the gamma matrices, specifically, $\gamma_0\gamma_5 = -\gamma_5\gamma_0$. The Lagrangian transforms under $\Lambda_A$ as follows

$$i\bar{\psi}\slashed{\partial}\psi \longrightarrow i\bar{\psi}\slashed{\partial}\psi - i\vec{\Theta}\left(\bar{\psi}\,i\partial_\mu\gamma^\mu\gamma_5\frac{\vec{\tau}}{2}\,\psi + \bar{\psi}\,\gamma_5\frac{\vec{\tau}}{2}i\partial_\mu\gamma^\mu\,\psi\right) \tag{43}$$

$$= i\bar{\psi}\slashed{\partial}\psi \tag{44}$$

where the second term vanishes because $\gamma_5$ anti-commutes with $\gamma_\mu$. Thus the Lagrangian is also invariant under $\Lambda_A$ with the conserved 'axial - vector' current

$$A_\mu^a = \bar{\psi}\gamma_\mu\gamma_5\frac{\tau^a}{2}\psi \tag{45}$$

In summary, the Lagrangian of massless fermions, and, hence, massless QCD, is invariant under both transformations, $\Lambda_V$ and $\Lambda_A$[3] symmetry is what is meant by chiral symmetry[4].

---

[3]Note, that the above Lagrangian is also invariant under the operations $\psi \rightarrow exp(-i\Theta)\psi$ and $\psi \rightarrow exp(-i\gamma_5\Theta)\psi$. The first operation is related to the conservation of the baryon number while the second symmetry is broken on the quantum level. This is referred to as the U(1) axial anomaly, which is real breaking of the symmetry in contrast to the spontaneous breaking discussed below (see e.g. [15]).

[4]Often, people talk about 'chiral' symmetry but actually only refer to the axial transformation $\Lambda_A$. This is due to its special role is plays, since it is spontaneously broken in the ground state.

The chiral symmetry is often referred to by its group structure as the $SU(2)_V \times SU(2)_A$ symmetry.

Now let us see, what happens if we introduce a mass term.

$$\delta \mathcal{L} = -m\left(\bar{\psi}\psi\right) \tag{46}$$

From the above, $\delta \mathcal{L}$ is obviously invariant under the vector transformations $\Lambda_V$ but *not* under $\Lambda_A$

$$\Lambda_A : \; m\left(\bar{\psi}\psi\right) \longrightarrow \bar{\psi}\psi - 2i\vec{\Theta}\left(\bar{\psi}\frac{\vec{\tau}}{2}\gamma_5\psi\right) \tag{47}$$

Thus, $\Lambda_A$ is not a good symmetry, if the fermions (quarks) have a finite mass. But as long as the masses are small compared to the relevant scale of the theory one may treat $\Lambda_A$ as an approximate symmetry, in the sense, that predictions based under the assumption of the symmetry should be reasonably close to the actual results[5].

In case of QCD we know that the masses of the light quarks are about $5 - 10\,\text{MeV}$ whereas the relevant energy scale given by $\Lambda_{QCD} \simeq 200\,\text{MeV}$ is considerably larger. We, therefore, expect that $\Lambda_A$ should be an approximate symmetry and that the axial current should be approximately (partially) conserved. This slight symmetry breaking due to the quark masses is the basis of the so called Partial Conserved Axial Current hypothesis (PCAC). Furthermore, as long as the symmetry breaking is small, one would also expect, that its effect can be described in a perturbative approach. This is carried out in a systematic fashion in the framework of chiral perturbation theory.

## 2.3 Chiral Symmetry and PCAC

### 2.3.1 Chiral transformation of mesons

In order to develop a better feeling for the meaning of the symmetry transformations $\Lambda_V$ and $\Lambda_A$, let us find out pions and rho-mesons transform under these operations. To this end, let us consider combinations of quark fields, which carry the quantum numbers of the mesons under consideration. This should give us the correct transformation properties:

pion-like state: $\vec{\pi} \equiv i\bar{\psi}\vec{\tau}\gamma_5\psi;$      sigma-like state:    $\sigma \equiv \bar{\psi}\psi$

rho-like state: $\vec{\rho}_\mu \equiv \bar{\psi}\vec{\tau}\gamma_\mu\psi;$      $a_1$-like state:    $\vec{a}_{1\,\mu} \equiv \bar{\psi}\vec{\tau}\gamma_\mu\gamma_5\psi$

---

[5] A wheel which is slightly bent and thus not perfectly invariant under rotations, can for most practical purposes still be considered as being round, as long as the bending is small compared to the radius of the wheel.

**(i)** vector transformations $\Lambda_V$, see eqs. (37,38):

$$\pi_i: \quad i\bar{\psi}\tau_i\gamma_5\psi \quad \longrightarrow \quad i\bar{\psi}\tau_i\gamma_5\psi + \Theta_j\left(\bar{\psi}\tau_i\gamma_5\frac{\tau_j}{2}\psi - \bar{\psi}\frac{\tau_j}{2}\tau_i\gamma_5\psi\right)$$
$$= i\bar{\psi}\tau_i\gamma_5\psi + i\Theta_j\epsilon_{ijk}\bar{\psi}\gamma_5\tau_k\psi \qquad (48)$$

where we have used the commutation relation between the $\tau$ matrices $[\tau_i, \tau_j] = 2i\epsilon_{ijk}\tau_k$. In terms of pions this can be written as

$$\vec{\pi} \longrightarrow \vec{\pi} + \vec{\Theta} \times \vec{\pi} \qquad (49)$$

which is nothing else than an isospin rotation, namely the isospin direction of the pion is rotated by $\Theta$. The same result one obtains for the $\rho$ - meson

$$\vec{\rho} \longrightarrow \vec{\rho} + \vec{\Theta} \times \vec{\rho} \qquad (50)$$

Consequently, the vector-transformation $\Lambda_V$ can be identified with the isospin rotations and the conserved vector current with the isospin current, which we know to be conserved in strong interactions.

**(i)** axial transformations $\Lambda_A$, see eqs. (41,42):

$$\pi_i: \quad i\bar{\psi}\tau_i\gamma_5\psi \quad \longrightarrow \quad i\bar{\psi}\tau_i\gamma_5\psi + \Theta_j\left(\bar{\psi}\tau_i\gamma_5\gamma_5\frac{\tau_j}{2}\psi + \bar{\psi}\gamma_5\frac{\tau_j}{2}\tau_i\gamma_5\psi\right)$$
$$= i\bar{\psi}\tau_i\gamma_5\psi + \Theta_i\bar{\psi}\psi \qquad (51)$$

where we have made use of the anti-commutation relation of the $\tau$ matrices $\{\tau_i, \tau_j\} = 2\delta_{ij}$ and of $\gamma_5\gamma_5 = 1$. In terms of the mesons this reads:

$$\vec{\pi} \longrightarrow \vec{\pi} + \vec{\Theta}\sigma \qquad (52)$$

The pion and the sigma-meson are obviously rotated into each other under the axial transformations $\Lambda_A$. Similarly the rho rotates into the $a_1$

$$\vec{\rho}_\mu \longrightarrow \vec{\rho}_\mu + \vec{\Theta} \times \vec{a_1}_\mu \qquad (53)$$

Above we just have convinced ourselves that $\Lambda_A$ is a symmetry of the QCD Hamiltonian. Naively, this would imply, that states which can be rotated into each other by this symmetry operation should have the same Eigenvalues, i.e the same masses. This, however, is clearly not the case, since $m_\rho = 770\,\text{MeV}$ and $m_{a_1} = 1260\,\text{MeV}$. We certainly do not expect that the slight symmetry breaking due to the finite current quark masses is responsible for this splitting. This should lead to mass differences which are small compared to the masses themselves. In case of the $\rho$ and $a_1$, however, the mass difference is of the same order as the mass of the $\rho$. The resolution to this problem will be the spontaneous breakdown of the axial symmetry. Before we discuss what is meant by that, let us first convince ourselves, that the axial vector is conserved to a good approximation, so that the axial symmetry must be present somehow.

### 2.3.2 Pion decay and PCAC

Let us first consider the weak decay of the pion. In the simple Fermi theory the weak interaction Hamiltonian is of the current-current type, where both currents are a sum of axial and vector currents, as we have defined them above (see e.g. [2]). Because of parity, the weak decay of the pion is controlled by the matrix element of the axial current between the vacuum and the pion $<0|A_\mu|\pi>$. This matrix element must be proportional to the pion momentum, because this is the only vector around

$$<0|A_\mu^a(x)|\pi^b(q)> = if_\pi q_\mu \delta^{ab} e^{-iq\cdot x} \tag{54}$$

and the proportionality constant $f_\pi = 93 MeV$ is determined from experiment[6]. Let us now take the divergence of equ. (54)

$$<0|\partial^\mu A_\mu^a(x)|\pi^b(q)> = -f_\pi q^2 \delta^{ab} e^{-iq\cdot x} = -f_\pi m_\pi^2 \delta^{ab} e^{-iq\cdot x} \tag{55}$$

To the extent, that the pion mass is small compared to hadronic scales, the axial current is approximately conserved. Or in other words, the smallness of the pion mass is directly related to the partial conservation of the axial current, i.e. to the fact that the axial transformation is an approximate symmetry of QCD. In the literature the above relation (55) is often referred to as the PCAC relation. The above relations (54,55) also suggest, that the axial current carried by a pion is

$$A_\mu^\pi = f_\pi \partial_\mu \Phi(x) \tag{56}$$

or that the divergence of the axial-vector current can be identified with the pion field (up to a constant). Here $\Phi(x)$ is the pion field. Sometimes this relation between pion field and axial current is also referred to as the PCAC relation.

### 2.3.3 Goldberger-Treiman relation

There is more evidence for the conservation of the axial current. Let us consider the axial current of a nucleon. This is simply given by (see equ. (45))

$$A_\mu^N = g_a \bar{\psi}_N \gamma_\mu \gamma_5 \frac{\tau}{2} \psi_N \tag{57}$$

where $\psi_N = (proton, neutron)$ is now an isospinor representing proton and neutron. The factor $g_a = 1.25$, is due to the fact, that the axial current of the nucleon is renormalized by 25%, as seen in the weak beta decay of the neutron. Since the nucleon has a large mass $M_N$, we do not expect that its axial current is conserved, and indeed by using the free Dirac equation for the nulceon one can show that

$$\partial^\mu A_\mu^N = i g_a M_N \bar{\psi}_N \gamma_5 \tau \psi_N \neq 0 \tag{58}$$

which vanishes only in case of vanishing nucleon mass. We know, however, that the nucleon interacts strongly with the pion. Therefore, let us assume that the total axial current is the sum of the nucleon and the pion contribution. Using the PCAC-relation (56) and equ. (57) we have

$$A_\mu = g_a \bar{\psi}_N \gamma_\mu \gamma_5 \frac{\tau}{2} \psi_N + f_\pi \partial_\mu \Phi(x) \tag{59}$$

---

[6]The are several definitions of $f_\pi$ around, depending on whether factors of 2, $\sqrt{2}$ are present in equ. (54).

If we require, that the total current is conserved, $\partial^\mu A_\mu = 0$, we obtain

$$\partial^\mu \partial_\mu \Phi = -g_a \, i \frac{M_N}{f_\pi} \bar{\psi}_N \gamma_5 \tau \psi_N \tag{60}$$

where we have used (58). This is nothing else but a Klein Gordon equation for a massless boson (pion) coupled to the nucleon. Hence, requiring the conservation of the total axial current immediately leads us to predict that the pion should be massless. This is exactly what we also concluded from the weak pion decay. If we now allow for a finite pion mass, which is equivalent to requiring that the divergence of the axial current is consistent with the PCAC result (55), then we arrive at the Klein Gordon equation for a pion coupled to the nucleon

$$\left(\partial^\mu \partial_\mu + m_\pi^2\right) \Phi = -g_a \, i \frac{M_N}{f_\pi} \bar{\psi}_N \gamma_5 \tau \psi_N \tag{61}$$

where the pion-nucleon coupling constant is given by

$$g_{\pi NN} = g_a \frac{M_N}{f_\pi} \simeq 12.5 \tag{62}$$

This is to be compared with the value for the pion-nucleon coupling as extracted e.g. from pion-nucleon scattering experiments

$$g_{\pi NN}^{exp} = 13.4 \tag{63}$$

which is in remarkeble close agreement, considering the fact, that equ. (62) relates the strong-interaction pion-nucleon coupling $g_{\pi NN}$ with quantities extracted from the weak interaction, namely $g_a$ and $f_\pi$. Of course, the reason why this works is that there is some symmetry, namely chiral symmetry, at play, which allows to connect semingly different pieces of physics. Equation (62) is usually called the Goldberger-Treiman relation.

### 2.3.4 Spontaneous breakdown of chiral symmetry

There appears to be some contradiction: On the one hand the meson mass sepctrum does not reflect the axial-vector symmetry. On the other hand, the weak pion decay seems to be consistent with a (partially) conserved axial-vector current. Also the success of the Goldberger-Treiman relation indicates that the axial-vector current is conserved and, hence, that the axial transformation $\Lambda_A$ is a symmetry of the strong interactions.

The solution to this puzzle is, that the axial-vector symmetry is *spontaneously* broken. What does one mean by that? One speaks of a spontaneously broken symmetry, if a symmetry of the Hamiltonian is not realized in the ground state.

This is best illustrated in a classical mechanics analog. In fig. 1 we have two rotationally invariant potentials ('interactions'). In (a) the ground state is right in the middle, and the potential plus ground state are still invariant under rotations. In (b), on the other hand, the ground state is at a finite distance away from the center. The point at the center is a local maximum of the potential and thus unstable. If we put a little ball in the middle, it will roll down somewhere and find its groundstate some place in the valley which represents the true minimum of the potential. By picking one point in this valley (i.e picking the ground state), the rotational symmetry is obviously broken. Potential plus groundstate are not symmetric anymore. The symmetry has been broken *spontaneously* by choosing a certain direction to be the groundstate. However, effects of the symmetry are still present. Moving the ball around in the valley (rotational excitations) does not cost any energy, whereas radial excitations do cost energy.

Let us now use this mechanics analogy in order to understand what the spontaneous breakdown of the axial-vector symmetry of the strong interaction means. Assume, that the effective QCD-hamiltonian at zero temperature has a form similar to that depicted in fig. 1(b), where the (x,y)-coordinates are replaced by $(\sigma, \vec{\pi})$-fields. The spacial rotations are then the mechanics analog of the axial-vector rotation $\Lambda_A$, which rotates $\vec{\pi}$ into $\sigma$ (see equ. (52)). Since the ground state is not in the center but a some finite distance away from it, one of the fields will have a finite expectation value. This can only be the $\sigma$-field, because it carries the quantum numbers of the vacuum. In the quark language, this means we expect to have a finite scalar quark condensate $< \bar{q}q > \neq 0$. In this picture, pionic excitation correspond to small 'rotations' away from the ground-state along the valley, which do not cost any energy. Consequently the mass of the pion should be zero. In other words, due to the spontaneous breakdown of chiral symmetry, we predict a vanishing pion mass. Excitations in the $\sigma$-direction correspond to radial excitations and therefore are massive.

This scenario is in perfect agreement with what we have found above. The spontaneous breakdown of the axial-vector symmetry leads to different masses of the pion and sigma. However, since the interaction itself is still symmetric, pions become massless, which is exactly what we find from the PCAC relation, provided that the axial current is perfectly conserved. Thus the mesonic mass spectrum as well as the PCAC– and the Goldberger-Treiman relation are consistent with a spontaneous breakdown of the axial-vector symmetry $\Lambda_A$. The pion appears as a massless mode (Goldstone boson) as a result of the symmetry of the interaction.

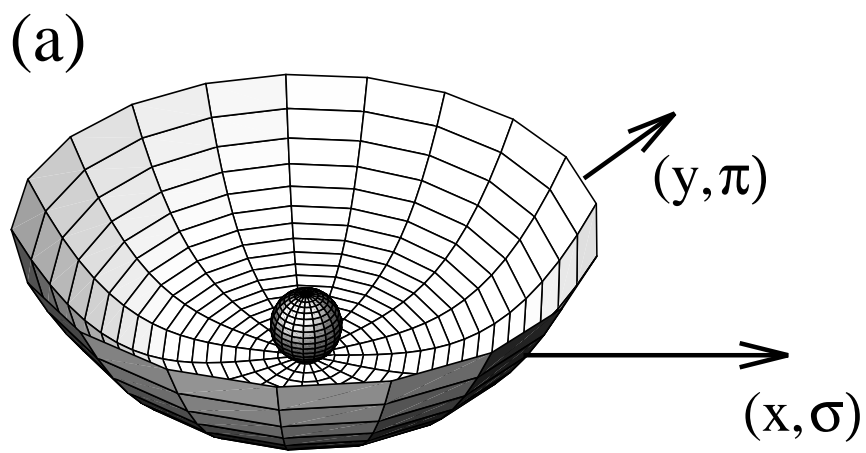

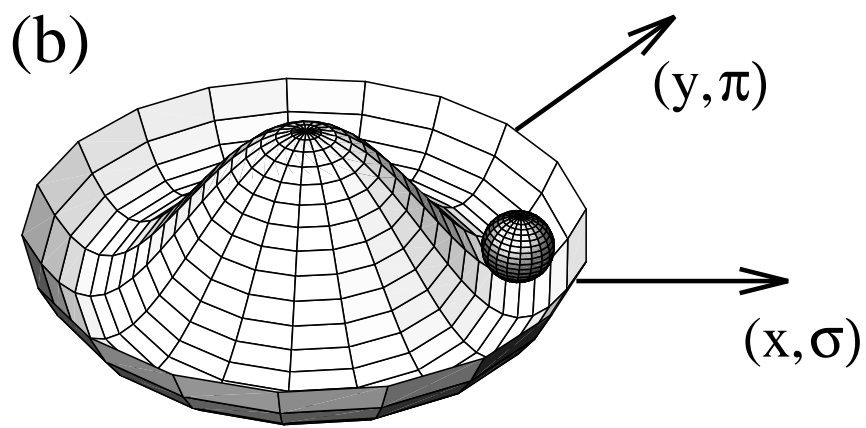

Figure 1: Effective potentials. (a) No spontaneous breaking of symmetry. (b) Spontaneous breaking of symmetry.

Incidentally, the assumption of a spontaneously broken axial-vector symmetry also explains the mass difference between the $\rho$- and $a_1$ meson and one predicts that $m_{a_1} = \sqrt{2} m_\rho$ in good agreement with the measured masses. The derivation of this result, however, is too involved to be presented here and the interested reader is referred to the literature [3, 4].

One expects, that at high temperature/densities the finite expectation value of the scalar quark condensate melts away resulting in a system, where chiral symmetry is not spontaneously broken anymore. In this, as it is often called, chirally restored phase pion/sigma as well as rho/$a_1$, if they exist[7], should be degenerate and the pion looses its identity as a Goldstone boson, i.e. it will become massive. The effective interaction in this phase would then have a shape similar to fig 1(a). It is one of the major goals of the ultrarelativistic heavy ion program to create and identify a macroscopic sample of this phase in the laboratory.

In the following section we will construct a chiral invariant Lagrangian, the so called 'Linear-sigma-model', in order to see how the concept of spontaneous breakdown of chiral symmetry is realized in the framework of a simple model. We will also discuss how to incoorporate the effect of the finite quark masses leading to the explicit breaking of chiral symmetry.

---

[7]If deconfinement and chiral restoration occur at the same temperature, it may become meaningless to talk about mesons above the critical temperature.

# 3  Linear sigma-model

## 3.1  Chiral limit

In this section we will construct a simple chirally invariant model involving pions and nucleons, the so called linear sigma - model. This model was first introduced by Gell-Mann and Levy in 1960 [5], long before QCD was known to be the theory of the strong interaction. In order to construct such a model, we have to write down a Lagrangian which is a Lorentz-scalar and which is invariant under the vector- and axial-vector transformations, $\Lambda_V$ and $\Lambda_A$.

In the previous section, we have shown, that the pion transforms under $\Lambda_V$ and $\Lambda_A$ as (52).

$$\Lambda_V : \pi_i \longrightarrow \pi_i + \epsilon_{ijk}\Theta_j\pi_k \qquad \Lambda_A : \pi_i \longrightarrow \pi_i + \Theta_i\sigma \tag{64}$$

Similarly one can also show, that the $\sigma$-field transforms like

$$\Lambda_V : \sigma \longrightarrow \sigma \qquad \Lambda_A : \sigma \longrightarrow \sigma - \Theta_i\pi_i \tag{65}$$

Since $\Lambda_V$ is simply an isospin rotation, the squares of the fields are invariant under this transformation

$$\Lambda_V : \quad \pi^2 \longrightarrow \pi^2; \qquad \sigma^2 \longrightarrow \sigma^2 \tag{66}$$

whereas under $\Lambda_A$ they transform like

$$\Lambda_A : \quad \pi^2 \longrightarrow \pi^2 - 2\sigma\Theta_i\pi_i; \qquad \sigma^2 \longrightarrow \sigma^2 + 2\sigma\Theta_i\pi_i \tag{67}$$

However, the combination $(\pi^2 + \sigma^2)$ is invariant under both transformations, $\Lambda_V$ *and* $\Lambda_A$

$$(\pi^2 + \sigma^2) \xrightarrow{\Lambda_V,\Lambda_A} (\pi^2 + \sigma^2) \tag{68}$$

Since this combination is also a Lorentz-scalar, we can build a chirally invariant Lagrangian around this structure:

- Pion-nucleon interaction:
  The standard pion nucleon interaction involves a pseudo-scalar combination of the nucleon field multiplied by the pion field:

$$g_\pi \left(i\bar\psi\gamma_5\vec\tau\psi\right)\vec\pi \tag{69}$$

  where from now on we denote the pion-nucleon coupling constant simply by $g_\pi$. Under the chiral transformations this transforms exactly like $\pi^2$, because the term involving the nucleon has the same quantum numbers as the pion. Chiral invariance requires that there must be another term, which transforms like $\sigma^2$, in order to have the invariant structure (68). The simplest choice is a term of the form,

$$g_\pi \left(\bar{\psi}\psi\right) \sigma \tag{70}$$

so that the interaction term between nucleons and the mesons is

$$\delta \mathcal{L} = -g_\pi \left[\left(\bar{\psi}\gamma_5\vec{\tau}\psi\right)\vec{\pi} + \left(\bar{\psi}\psi\right)\sigma\right] \tag{71}$$

- Nucleon mass term:
  We know that an explicit nucleon mass term breaks chiral invariance (see section 2.2.1 ). The nucleon mass is also too large to be simply a result of the small explicit chiral symmetry breaking as reflected in the PCAC relation (55). The simplest[8] way to give the nucleon a mass without breaking chiral symmetry, is to exploit the coupling of the nucleon to the $\sigma$-field (70), which has the structure of a nucleon mass term. This, however, requires that the $\sigma$-field as a *finite vacuum expectation value*,

$$< \sigma > = \sigma_0 = f_\pi \tag{72}$$

where choice of $\sigma_0 = f_\pi$ is dictated by the Goldberger-Treiman relation (62) in the limit of $g_a = 1$. A finite vacuum expectation value for the $\sigma$-field immediately implies, that chiral symmetry will be spontaneously broken, as discussed in the last section. In order for our model to generate such an expectation value, we have to introduce a potential for the sigma field, which has its minimum at $\sigma = f_\pi$. This brings us to the next ingredient of our model.

- Pion - sigma potential:
  The potential, which generates the vacuum expectation value of the $\sigma$ field has to be a function of the invariant structure (68) in order to be chirally invariant. The simplest choice is:

$$V = V(\pi^2 + \sigma^2) = \frac{\lambda}{4}\left(\left(\pi^2 + \sigma^2\right) - f_\pi\right)^2 \tag{73}$$

This potential, which is plotted in fig. (2) (see also fig. (1) for a three-dimensional view ) indeed has its minimum at $\sigma = f_\pi$ for $\pi = 0$. Due to its shape, it is often referred to as the 'Mexican - hat - potential'.

---

[8]Actually one can allow for an explicit nucleon mass term if one also includes the chiral partner of the nucleon, which is believed to be the $N^*(1535)$. This is an interesting alternative approach which is discussed in detail in ref. [6]

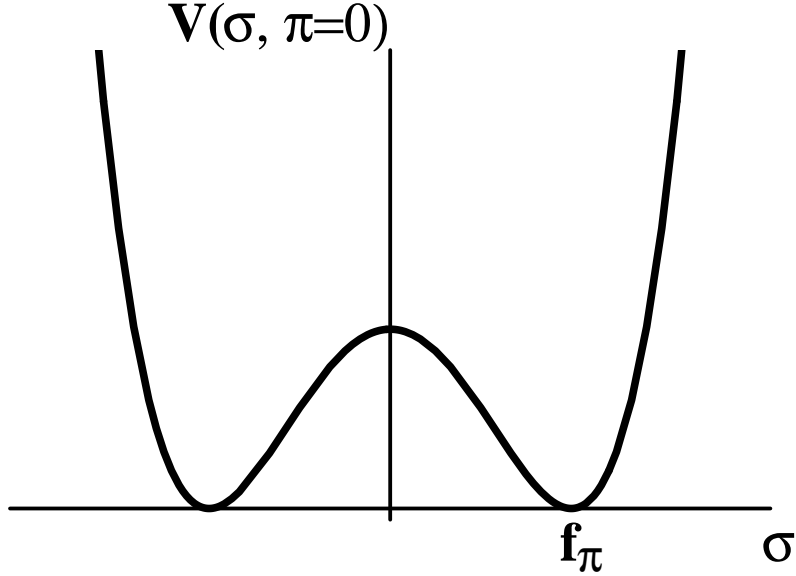

Figure 2: Potential of linear sigma-model

- Kinetic energy terms:
  Finally we have to add kinetic energy terms for the nucleons and the mesons which have the form $i\bar{\psi}\slashed{\partial}\psi$ and $\frac{1}{2}(\partial_\mu \pi \partial^\mu \pi + \partial_\mu \sigma \partial^\mu \sigma)$, respectively. Both are chirally invariant. The first term is just the Lagrangian of free mass less fermions, which we have shown to be invariant. The second term again has the invariant structure (68).

Putting everything together, the Lagrangian of the linear sigma-model reads (remember that the potential V enters with a minus-sign into the Lagrangian):

$$\mathcal{L}_{L.S.} = i\bar{\psi}\slashed{\partial}\psi - g_\pi\left(\bar{\psi}\gamma_5\vec{\tau}\psi\,\vec{\pi} + \bar{\psi}\psi\,\sigma\right) \\ -\frac{\lambda}{4}\left((\pi^2+\sigma^2) - f_\pi^2\right)^2 + \frac{1}{2}\partial_\mu\pi\partial^\mu\pi + \frac{1}{2}\partial_\mu\sigma\partial^\mu\sigma \qquad (74)$$

What are the properties of this model? Let us start with the ground state. As already mentioned, in the ground state the $\sigma$ - field has a finite expectation value, whereas the pion has none, because of parity. Furthermore, the nucleon obtains its mass from its interaction with the sigma field. But what are the masses of the $\sigma$ and $\pi$ - mesons? There are no explicit mass terms for the $\sigma$- and $\pi$-fields in the Lagrangian (74), but, as with the nucleon, there could be some coupling to the expectation value of the $\sigma$ field, which gives rise to mass terms. From the structure of the potential (see figs (2) and (1)) as well as from our discussion of the spontaneous breakdown of chiral symmetry, we expect the pion to be massless and the $\sigma$-meson to become massive. In order to verify that, let us expand the potential (73) for small fluctuations around the ground state.

$$\sigma = \sigma_0 + (\delta\sigma); \quad \pi = (\delta\pi) \tag{75}$$

Actually, it is these fluctuations $((\delta\sigma), (\delta\pi))$, which are to be be identified with the observed particles ($\sigma$- and $\pi$- meson). Since a bosonic mass term is quadratic in the fields (see Lagrangian (15)), let us expand the potential up to quadratic order in the fluctuations $(\delta\sigma)(\delta\pi)$. Expanding around a minimum, the linear order vanishes, and we have:

$$V(\sigma, \pi) = \lambda f_\pi^2 (\delta\sigma)^2 + \mathcal{O}(\delta^3) \tag{76}$$

where we have used that $\sigma_0 = f_\pi$. Comparing with the Lagrangian of a free boson we identify the mass of the sigma to be (remember that $L = T - V$)

$$m_\sigma^2 = \lambda f_\pi^2 \neq 0 \tag{77}$$

We find no mass term for the pion in agreement with our expectation, that the pion should be the massless Goldstone boson of the spontaneously broken chiral symmetry.

In summary, the properties of the ground state of the linear sigma-model are:

$$<\sigma> = \sigma_0 = f_\pi \tag{78}$$
$$<\pi> = 0 \tag{79}$$
$$M_N = g_\pi \sigma_0 = g_\pi f_\pi \tag{80}$$
$$m_\sigma^2 = \lambda f_\pi^2 \neq 0 \tag{81}$$
$$m_\pi = 0 \tag{82}$$

Before we conclude this section, let us calculate the conserved axial current and check, if the PCAC-relation is satisfied in our model. The infinitesimal axial transformations of the nucleon, pion and sigma fields are given by (see (41), (64) and (65))

$$\psi \longrightarrow \psi - i\gamma_5 \frac{\tau^a}{2}\Theta^a \psi \tag{83}$$
$$\pi^i \longrightarrow \pi^i + \Theta^a \delta^{i,a} \sigma \tag{84}$$
$$\sigma \longrightarrow \sigma - \Theta^a \pi^a \tag{85}$$

Comparing with the general form (29) for unitary transformations, we find that the generator of the axial transformation $T^a$ act on the fields in the following way

$$T^a \psi = \gamma_5 \frac{\tau^a}{2} \Theta^a \psi \qquad (86)$$

$$T^a \pi^j = i\sigma \delta^{a,j} \qquad (87)$$

$$T^a \sigma = -i\pi^a \qquad (88)$$

Using the expression for the conserved current (31) the conserved axial current is given by

$$A_\mu^a = \bar{\psi}\gamma_\mu \frac{\tau^a}{2}\psi - \pi^a \partial_\mu \sigma + \sigma \partial_\mu \pi^a \qquad (89)$$

In order to check the PCAC-relation, we again expand the fields around the ground state (see eq. (75))

$$A_\mu^a = \bar{\psi}\gamma_\mu \frac{\tau^a}{2}\psi - (\delta\pi^a)\partial_\mu(\delta\sigma) + (\delta\sigma)\partial_\mu(\delta\pi^a) + f_\pi \partial_\mu(\delta\pi^a) \qquad (90)$$

where we have used that $\sigma_0 = f_\pi$. Since the PCAC-relation involves the matrix element $<0|A_\mu^a|\pi^j>$ only the last term of (90) contributes. The other terms would require either nucleons or sigma-mesons in the final or initial state. Thus, as far as the PCAC relation is concerned, the axial current reduces to $((\delta\pi) = \pi)$

$$A_\mu^a(x)_{PCAC} = f_\pi \partial_\mu \pi(x) \qquad (91)$$

in agreement with the PCAC-results eq. (56).

## 3.2 Explicit breaking of chiral symmetry

So far we have assumed that the axial-vector symmetry is a perfect symmetry of the strong interactions. From our discussion in section 2.2.1 we know, however, that the small but finite current quark masses of the up and down quark break the axial-vector symmetry explicitly. This explicit breaking of the symmetry should not be confused with the spontaneous breakdown, we have discussed before. In case of a spontaneous breaking of a symmetry the Hamiltonian is still symmetric, whereas in case of an explicit breaking, already the Hamiltonian is not symmetric.

One may wonder if the whole concept of spontaneous symmetry breaking makes any sense if already the Hamiltonian is not symmetric. The answer to that, again, depends on the scales involved. If the explicit symmetry breaking is small, i.e. if the quark masses are small compared to to relevant energy scale of QCD, as we believe they are, then it will be sensible to apply the notion of a spontaneously broken symmetry.

To illustrate that, let us again utilize our little mechanics analogy, which we have developed in the previous section. An explicit symmetry breaking would imply that both potentials of figure (1) are not invariant under rotation. This could for instance be achieved by slightly tilting them towards, say, the x-direction. As a result, also the ground state of potential (a) is away from the center $(x, y = 0)$. But the dislocation is small compared to that due to the spontaneous breaking. Furthermore, as long as the potentials are tilted only slightly, rotational excitation (pions) in potential (b) are still considerably softer than the radial ones (sigma-mesons). So in this sense, we expect the effect due to the spontaneous breakdown of chiral symmetry to dominate the dynamics, as long as the explicit breaking is small. In the linear sigma-model, the mass scale generated by the spontaneous breakdown is the nucleon mass, whereas that generated by the explicit breakdown will be the mass of the pion, as we shall see. Thus, indeed the explicit breaking is small, and our picture, developed under the assumption of perfect axial-vector symmetry, will survive the introduction of the explicit breaking to a very good approximation.

After these remarks let us now introduce a symmetry breaking term into the linear sigma-model. In QCD, we know, that the symmetry is explicitly broken by a quark mass-term

$$\delta \mathcal{L}_{X\chi SB} = -m\bar{q}q \qquad (92)$$

where the subscript $X\chi SB$ stands for explicit chiral symmetry breaking. If we identify, as we have done before, the scalar quark-field combination $\bar{q}q$ with the $\sigma$ field, this would suggest the following symmetry breaking term in the sigma-model

$$\delta \mathcal{L}_{SB} = \epsilon \sigma \qquad (93)$$

where $\epsilon$ is the symmetry breaking parameter. This term clearly is not invariant under the axial transformation $\Lambda_A$ but preserves the vector symmetry $\Lambda_V$. Including this term, the potential $V$ (73) now has the form

$$V(\sigma, \pi) = \frac{\lambda}{4}\left((\pi^2 + \sigma^2) - v_0\right)^2 - \epsilon \sigma \qquad (94)$$

where we now have replaced $f_\pi$ of eq. (73) by a general parameter $v_0$, which in limit of $\epsilon \to 0$ will go to $f_\pi$. The effect of the symmetry breaking term is to tilt the potential slightly towards the positive $\sigma$ direction, and thus to break the symmetry (see fig. (3)).

What are the consequences of this additional term? First of all, the minimum has shifted slightly. If we require that the value of the new minimum is still $f_\pi$ in order to preserve the Goldberger-Treiman relation, we find for the parameter $v_0$ to leading order in $\epsilon$

$$v_0 = f_\pi - \frac{\epsilon}{2\lambda f_\pi^2} \qquad (95)$$

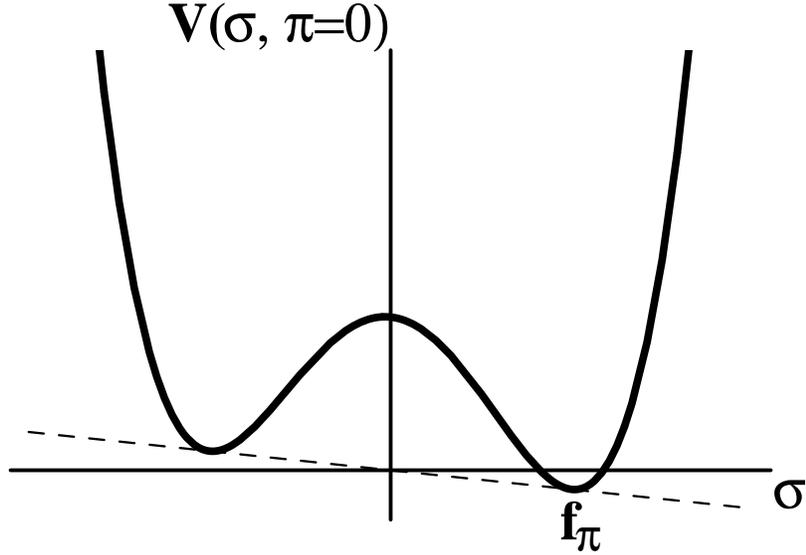

Figure 3: Potential of linear sigma-model with explicit symmetry breaking

Also the mass of the sigma is slightly changed

$$m_\sigma^2 = \left.\frac{\partial^2 V}{\partial \sigma^2}\right|_{\sigma_0} = 2\lambda f_\pi + \frac{\epsilon}{f_\pi} \qquad (96)$$

But most importantly, the pion now acquires a finite mass

$$m_\pi^2 = \left.\frac{\partial^2 V}{\partial \pi^2}\right|_{\sigma_0} = \frac{\epsilon}{f_\pi} \neq 0 \qquad (97)$$

which fixes the parameter $\epsilon$

$$\epsilon = f_\pi m_\pi^2 \qquad (98)$$

Thus, the square of the pion mass is directly proportional to the symmetry breaking parameter $\epsilon$ as we would have expected it from our previous discussion.

Due to our choice of $\sigma_0 = f_\pi$, the nucleon mass is not changed, which, however, does not mean that there is no contribution to the nucleon mass from the explicit symmetry breaking. If we split the nucleon mass into a contribution from the symmetric part of the potential ($\sim v_0$) and one from the symmetry breaking term ($\sim \epsilon$),

$$M_N = g_\pi \sigma_0 = g_\pi \left(v_0 + \frac{\epsilon}{2\lambda f_\pi^2}\right) \tag{99}$$

we find that the contribution from the symmetry breaking, which is often referred to as the pion-nucleon sigma-term[9], is given by

$$\Sigma_{\pi N} = \delta M_N^{X\chi SB} = g_\pi \frac{\epsilon}{2\lambda f_\pi^2} \simeq g_\pi f_\pi \frac{m_\pi^2}{m_\sigma^2} \tag{100}$$

As we shall see below, the pion-nucleon sigma-term can be measured in pion-nucleon scattering experiments and its is currently believed to be [7] $\Sigma_{\pi N}(0) = 35 \pm 5$MeV.

Since chiral symmetry is now explicitly broken, the axial-vector current is not conserved anymore. The functional form of the axial current is the same, however, as in the symmetric case, eq. (89), because the symmetry breaking term (93) does not involve any derivatives (see equ. (31)). Its divergence is related to the variation of the symmetry breaking term in the Lagrangian, as shown at the end of section 2.2.

$$\partial^\mu A_\mu^a = \epsilon \, \delta(\sigma) = -f_\pi m_\pi^2 \pi^a \tag{101}$$

which leads directly to the PCAC relation (55). Here $\delta(\sigma)$ denotes the variation of the $\sigma$-field with respect to the axial-vector transformation $\Lambda_A$, not the fluctuation around the ground state. As in equ. (31) the angel $\Theta^a$ has been divided out.

The main effect of the explicit chiral symmetry breaking was to give the pion a mass. But we can utilize the symmetry breaking further to derive[10] some rather useful relations between expectation values of the scalar quark operator $\bar{q}q$ and measurable quantities like $f_\pi$, $m_\pi$, and $\Sigma_{\pi N}$.

When we introduced the symmetry breaking term into our model, we had required that it has the same transformation properties under the chiral transformations as the QCD-symmetry breaking term. The overall strength of the symmetry breaking, $\epsilon$ we then adjusted to reproduce the ground state properties, namely the pion mass. Therefore, it seems reasonable to expect, that that the vacuum expectation value of the symmetry breaking terms in QCD (92) and in the effective model (93) are the same.

$$<0|\,\epsilon\sigma\,|0> \;=\; <0|-m\bar{q}q|0> \tag{102}$$

---

[9]This definition of the pion-nucleon sigma term should be taken with some care. For a rigorous definition see e.g. [7, 8]. In the framework of the sigma-model, this definition, however, is correct to leading order in $\epsilon$.

[10]These 'derivations' are merely heuristic, but I feel they nicely demonstrate the physics which is going on. For a rigorous derivation see e.g. [8].

If we insert for $\epsilon = m_\pi^2 f_\pi$ and use $<0|\sigma|0> = f_\pi$ we arrive at the so called Gell-Mann – Oakes – Renner (GOR) relation [9, 8]

$$m_\pi^2 f_\pi^2 = -\frac{m_u + m_d}{2} <0|\bar{u}u + \bar{d}d|0> \quad (103)$$

where we have written out explicitly the average quark mass, $m$, and the quark operator $\bar{q}q$. The GOR relation is extremely useful, since it relates the quark condensate with $f_\pi$ and/or the pion mass with the current-quark mass.

Similarly, but less convincingly, one can argue, that the contribution to the nucleon mass due to chiral symmetry breaking, $\Sigma_{\pi N}$, is the expectation value of the symmetry breaking Hamiltonian $\delta H_{X\chi SB} = -\delta\mathcal{L}_{X\chi SB}$ between nucleon states. This leads to the exact expression of the pion-nucleon sigma-term in terms of QCD variables [7, 10]

$$\Sigma_{\pi N} = \frac{m_u + m_d}{2} <N|\bar{u}u + \bar{d}d|N> \quad (104)$$

This relation will turn out to be very helpful in order to estimate the change of the chiral condensate in nuclear matter at finite density.

## 3.3 S-wave pion-nucleon scattering

In order to see how chiral symmetry affects the dynamics, let us, as an example, study pion-nucleon scattering in the sigma-model. Let us begin by introducing some notation.

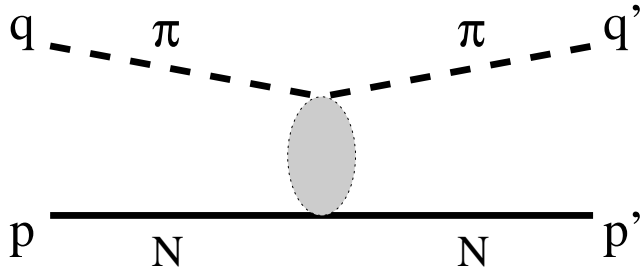

The invariant scattering amplitude $T(q, q')$ is commonly decomposed into a scalar and a vector part[11] (see fig. (3.3) for the notation of momenta)

$$T(q, q') = A(s,t) + \frac{1}{2}\gamma^\mu(q_\mu + q'_\mu)B(s,t) \qquad (105)$$

where $(s,t)$ are the usual Mandelstam variables, and $q$ and $q'$ denote the incoming and outgoing pion - four-momenta. The relativistic scattering amplitude is related to the more familiar scattering amplitude in the center of mass frame, $\mathcal{F}(\vec{q}, \vec{q}')$ by

$$\chi^+ \mathcal{F} \chi' = \frac{M_N}{4\pi\sqrt{s}} \bar{u}(p,s) T u(p',s') \qquad (106)$$

Here $\chi$ are Pauli-spinors for the nucleon representing spin and isospin and $u(p,s)$ stand for a relativistic spinors for a nucleon of momentum $p$.

The scattering amplitude can be decomposed into isospin-even and -odd components[12]

$$T_{ab} = T^+ \delta_{ab} + \frac{1}{2}[\tau_a, \tau_b] T^- \qquad (107)$$

where the indices $a, b$ refer to the isospin.

In the discussion of pion-nucleon scattering instead of (s,t) one usually uses the invariant variables [7]

$$\nu = \frac{s-u}{4M_N} \qquad (108)$$

$$\nu_B = -\frac{1}{2M_N} q^\mu q'_\mu = \frac{1}{4M_N}(t - q^2 - q'^2) \qquad (109)$$

---

[11] For details see e.g. the appendix of [11].

[12] Notice, that the isospin-odd amplitude is the *negative* of what in the literature is commonly called the iso-vector amplitude whereas the isospin-even amplitude is identical to the so called isoscalar one (see [11]).

The spin-averaged, non-spin-flip ($s = s'$), forward scattering ($p = p'$) amplitude, which will be most relevant for the aspects of chiral symmetry, is usually denoted by $D$ and is given in terms of the above variables by

$$D \equiv \frac{1}{2} \sum_s \bar{u}(p,s) T u(p,s) = A + \nu B \tag{110}$$

Finally, if one wants to extract effects due to explicit chiral symmetry breaking, one best analyses the so called subtracted amplitude

$$\bar{D} = D - D_{PV} = D - \frac{g_\pi^2}{M_N} \frac{\nu_B^2}{\nu_B^2 - \nu^2} \tag{111}$$

Now let us calculate the pion-nucleon scattering amplitude in the sigma- model. At tree level the diagrams shown in fig. (4) contribute to the amplitude. The first two processes represent the the simple absorption and re-emission of the pion by the nucleon. Provided, that there is a coupling between pion and nucleon, one would have written down these diagrams immediately, without any knowledge of chiral symmetry. The third diagram (c), which involves the exchange of a sigma-meson, is a direct result of chiral symmetry, and, as well shall see, is crucial in order to give the correct value for the amplitude.

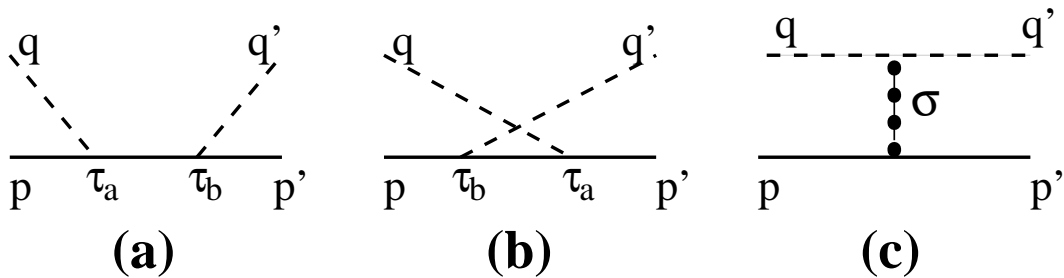

Figure 4: Diagrams contributing to the pion-nucleon scattering amplitude $T_{ab}$.

In the following, we will restrict ourselves to the forward scattering amplitudes, i.e. $q = q'$ and $p = p'$. Using standard Feynman-rules (see e.g. [1]), the above diagrams can be evaluated in a straightforward fashion. For diagram (a) we obtain

$$\bar{u}(p) T_{ab}^{(a)} u(p)$$
$$= g_\pi^2 \bar{u}(p) \tau_a \gamma_5 \frac{(p+q)^\mu \gamma_\mu + m}{(p+q)^2 - m^2} \tau_b \gamma_5 u(p)$$
$$= \bar{u}(p) \left[ (\delta_{ab} + \frac{1}{2}[\tau_a, \tau_b])(-g_\pi^2 \frac{q^\mu \gamma_\mu}{s - m^2}) \right] \tag{112}$$

where we have used that $\gamma_5\gamma_\mu = -\gamma_\mu\gamma_5$, $\gamma_5^2 = 1$, $\tau_a\tau_b = \delta_{ab} + \frac{1}{2}[\tau_a, \tau_b]$, and the Dirac equation $(p_\mu\gamma^\mu - m)u(p) = 0$. Obviously, diagram (a) contributes only to the vector piece of the amplitude, $B$, and the isospin-even and -odd amplitudes are the same

$$B^+_{(a)} = B^-_{(a)} = -\frac{g_\pi^2}{s - M_N^2} \tag{113}$$

The contribution of the crossed or u-channel ( diagram (b)) one obtains by replacing

$$s \rightarrow u \tag{114}$$
$$(\tau_a\tau_b) \rightarrow (\tau_b\tau_a) \tag{115}$$
$$q \rightarrow -q \tag{116}$$

with the result

$$B^+_{(b)} = \frac{g_\pi^2}{u - M_N^2} = -B^-_{(b)} \tag{117}$$

Here isospin-even and -odd amplitudes have the opposite sign.

It is instructive to calculate the scattering amplitude resulting from the first two diagrams only. If we didn't know about chiral symmetry, and, hence, the existence of the $\sigma$-exchange diagram, this is what we would naively obtain. At threshold ($\vec{q} = 0$), the combined amplitudes are

$$\nu B^+_{(a)+(b)} = -\frac{g_\pi^2}{M_N}\left(\frac{1}{1 - \frac{m_\pi^2}{4M_N^2}}\right) \tag{118}$$

$$\nu B^-_{(a)+(b)} = g_\pi^2\frac{m_\pi}{2M_N^2}\left(\frac{1}{1 - \frac{m_\pi^2}{4M_N^2}}\right) \tag{119}$$

Using equations (105, 106) the resulting s-wave isospin-even and isospin-odd scattering scattering length, which is related to the scattering amplitude D (110) at threshold by

$$a^\pm = \frac{1}{4\pi(1 + \frac{m_\pi}{M_N})}D^\pm_{\text{at threshold}} \tag{120}$$

would be

$$a_0^+((a) + (b)) = -\frac{g_\pi}{4\pi f_\pi(1 + \frac{m_\pi}{M_N})}\left(1 + \mathcal{O}(\frac{m_\pi^2}{M_N^2})\right) \simeq -1.4\,m_\pi^{-1} \tag{121}$$

$$a_0^-((a) + (b)) = \frac{m_\pi}{8\pi f_\pi^2(1 + \frac{m_\pi}{M_N})}\left(1 + \mathcal{O}(\frac{m_\pi^2}{M_N^2})\right) \simeq 0.078\,m_\pi^{-1} \tag{122}$$

where we have made of the Goldberger-Treiman relation $g_\pi f_\pi = M_N$. This is to be compared with the experimental values of [11]

$$a_0^+(exp) = -0.010(3) \, m_\pi^{-1} \qquad a_0^-(exp) = 0.091(2) \, m_\pi^{-1} \tag{123}$$

While we find reasonable agreement for the isospin-odd amplitude, the isospin even amplitude is off by two orders of magnitude! A different choice of the pion-nucleon coupling $g_\pi$ would not fix the problem, but just shift it from one amplitude to the other. Before we evaluate the remaining diagram (c), let us point out that in the chiral limit, i.e. $m_\pi = 0$, the isospin-odd amplitude vanishes.

In order to evaluate the $\sigma$-exchange diagram, we need to extract the pion-sigma coupling from our Lagrangian. This is done by expanding the potential $V$ (73) up to third power in the field fluctuations ($(\delta\pi)$ and $(\delta\sigma)$). The terms proportional to $\sim (\delta\pi)^2(\delta\sigma)$ then give the desired coupling.

$$\delta\mathcal{L}_{\pi\pi\sigma} = -\lambda f_\pi (\delta\pi)^2 (\delta\sigma) \tag{124}$$

The resulting amplitude is then given by

$$\bar{u}(p) T_{ab}^{(c)} u(p) = -g_\pi \frac{2\lambda f_\pi}{t - m_\sigma^2} \delta_{ab} \tag{125}$$

It only contributes to the scalar part of the amplitude, A, and only in the isospin-even channel. Using $2\lambda f_\pi^2 = m_\sigma^2 - m_\pi^2$ (see eqs. (96, 97) ) we find

$$A_{(c)}^+ = -\frac{g_\pi}{f_\pi} \frac{m_\sigma^2 - m_\pi^2}{m_\sigma^2 - t} = \frac{g_\pi}{f_\pi}\left(1 - \frac{t - m_\pi^2}{t - m_\sigma^2}\right) \tag{126}$$

To leading order, the contribution to the s-wave scattering lengths of diagram (c) is

$$a_0^+((c)) = \frac{g_\pi}{4\pi f_\pi(1 + \frac{m_\pi}{M_N})}(1 + \mathcal{O}(\frac{m_\pi^2}{M_N^2})) \tag{127}$$

$$a_0^-((c)) = 0 \tag{128}$$

Thus, to leading order, the contribution of the $\sigma$-exchange diagram (c) *exactly* cancels that of the nucleon-pole diagrams ((a) and (b)) and the total isospin-even scattering length vanishes

$$a_0^+ = 0 + \mathcal{O}(\frac{m_\pi^2}{M_N^2}, \frac{m_\pi^2}{m_\sigma^2}) \tag{129}$$

in much better agreement with experiment. The cancelation between the large individual contributions to the isospin-even amplitude is a direct consequence of chiral symmetry, which required the $\sigma$-exchange diagram. In the chiral limit, this cancelation is perfect, i.e. the isospin-even scattering amplitude vanishes identically, because the corrections $\sim m_\pi$ are zero in this case.

Furthermore, since the third diagram (c) does not contribute to the isospin-odd amplitude, the good agreement found above still holds. In other words, with the 'help' of chiral symmetry both amplitudes are reproduced well.

Putting all terms together the isospin-even amplitude $D^+$ is given in terms of the variables $\nu$ and $\nu_B$

$$\begin{aligned} D^+(\nu, \nu_B) &= A^+ + \nu B^+ \\ &= \frac{g_\pi}{f_\pi} \frac{\nu^2}{\nu_B^2 - \nu^2} + \frac{g_\pi}{f_\pi}\left(1 - \frac{t - m_\pi^2}{t - m_\sigma^2}\right) \\ &= \frac{g_\pi}{f_\pi} \frac{\nu_B^2}{\nu_B^2 - \nu^2} - \frac{g_\pi}{f_\pi} \frac{t - m_\pi^2}{t - m_\sigma^2} \end{aligned} \quad (130)$$

Here the first term in the second line is the contribution form diagrams (a) and (b) and the other two term are from diagram (c). At threshold, where $\nu = m_\pi$, $\nu_B = -\frac{m_\pi^2}{2M_N}$, and $t = 0$ this reduces to

$$D^+_{\text{at threshold}} = -\frac{g_\pi}{f_\pi}\left(\frac{m_\pi^2}{4M_N^2 - m_\pi^2} + \frac{m_\pi^2}{m_\sigma^2}\right) \quad (131)$$

As already pointed out, to leading order ($\sim m_\pi^0$) or in the chiral limit, this amplitude vanishes, as a result chiral symmetry. However the contribution next to leading order $\sim m_\pi^2$ involve also the mass of the $\sigma$-meson, which has not yet been clearly identified in experiment. In the sigma-model, this mass essentially is a free parameter, since it is directly proportional to the coupling $\lambda$. Since $\lambda$ gives the strength of the invariant potential $V$, chiral symmetry considerations will not determine this parameter. Thus, aside from the very important finding, that the isospin-even scattering length should be small, the linear sigma-model as no predictive power for the *actual* small value of the scattering length[13]

Notice, that although D is the spin averaged, forward ($t = 0$) scattering amplitude, we can obviously study it an any value of $\nu$, $t$ or equivalently $\nu$ and $\nu_B$. A kinematical point of particular interest is the so called Cheng-Dashen point, given by

$$\nu = 0, \qquad t = 2m_\pi \to \nu_B = 0 \quad (132)$$

---

[13]In the framework of chiral perturbation theory, the value of the isospin-even amplitude is essentially regarded as an input to fix the parameters of the expansion. There are attempts to relate the value of the scattering length to contributions from the Delta [12]. In this approach, the problem is shifted to the determination of an unknown off-shell parameter appearing in the Delta-propagator.

At this kinematical point, the subtracted amplitude $\bar{D}$ (111) is directly related with the pion-nucleon sigma-term $\Sigma_{\pi N}$ [7]

$$\bar{D}(\nu = 0, t = 2m_\pi) = \frac{\Sigma_{\pi N}}{f_\pi^2} \tag{133}$$

In the sigma-model we find for the subtracted amplitude to leading order in the pion mass

$$\bar{D}(\nu = 0, t = 2m_\pi) = -\frac{g_\pi}{f_\pi}\frac{m_\pi^2}{m_\sigma^2} = \frac{\Sigma_{\pi N}}{f_\pi^2} \tag{134}$$

where we have used the expression for the sigma-term, derived above (100) from the contribution of the explicit chiral symmetry breaking to the nucleon mass. Notice, although the Cheng-Dashen point is in an unphysical region, it can nevertheless be reached via dispersion relation techniques, and, thus, the sigma-term can be extracted from pion-nucleon scattering data. For a detailed discussion, see ref. [7].

# 4 Nonlinear sigma-model

One of the disturbing features of the linear sigma-model is the existence of the $\sigma$-field, because it cannot really be identified with any existing particle. Furthermore, at low energies and temperatures one would expect that excitations in the $\sigma$-direction should be much smaller than pionic ones, which in the chiral limit are massless (see fig. (1)). This is supported by our results for the pion-nucleon scattering, where in the final result the mass of the sigma-meson only showed up in next to leading order corrections, which vanish in the chiral limit.

Let us, therefore, remove the $\sigma$-meson as a dynamical field by sending its mass to infinity. Formally this can be achieved by assuming an infinitely large coupling $\lambda$ in the linear sigma-model. As a consequence the mexican hat potential gets infinitely steep in the sigma-direction (see fig. (4) ). This confines the dynamics to the circle, defined by the minimum of the potential.

$$\sigma^2 + \pi^2 = f_\pi^2 \tag{135}$$

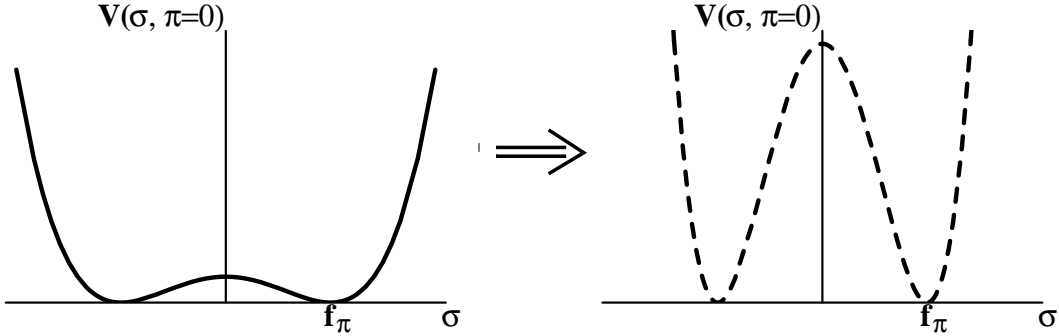

This additional condition removes one degree of freedom, which close to the ground state, where $<\sigma> = f_\pi$, is the sigma field, and we are left with pionic excitations only. Because of the above constraint (135), the dynamics is now restricted to rotation on the so called chiral circle (actually it is a sphere). Therefore, the fields can be expressed in terms of angles $\vec{\Phi}$,

$$\begin{aligned}
\sigma(x) &= f_\pi \cos(\frac{\Phi(x)}{f_\pi}) = f_\pi + \mathcal{O}(\Phi^2) \\
\vec{\pi}(x) &= f_\pi \hat{\Phi} \sin(\frac{\Phi(x)}{f_\pi}) = \vec{\Phi}(x) + \mathcal{O}(\Phi^3)
\end{aligned} \tag{136}$$

which to leading order can be identified with the pion field. Here, $\Phi = \sqrt{\vec{\Phi}\vec{\Phi}}$. Clearly, this ansatz fulfills the constraint (135). Equivalently, one can chose a complex notation for the fields, as it is commonly done in the literature

$$U(x) = e^{i\frac{\vec{\tau}\vec{\Phi}(x)}{f_\pi}} = \cos(\frac{\Phi(x)}{f_\pi}) + i\vec{\tau}\hat{\Phi}\sin(\frac{\Phi(x)}{f_\pi}) = \frac{1}{f_\pi}(\sigma + i\vec{\tau}\vec{\pi}) \tag{137}$$

where $U$ represents a unitary $(2 \times 2)$ matrix. The constraint (135) is then equivalent to

$$\frac{1}{2}tr(U^+ U) = \frac{1}{f_\pi}(\sigma^2 + \pi^2) = 1 \tag{138}$$

Since chiral symmetry, or more precisely axial-vector symmetry, corresponds to a symmetry with respect to rotation around the chiral circle, all structures of the form

$$tr(U^+U), \ tr(\partial_\mu U^+ \partial^\mu U) \ldots \tag{139}$$

are invariant. Already at this point it becomes obvious that we eventually will need some scheme, which tells us which structures to include and which ones not. This will lead us to the ideas of chiral perturbation theory in the following section.

Let us continue by rewriting the Lagrangian of the linear sigma-model (74) in terms of the new variables $U$ or $\Phi$. After a little algebra we find that the kinetic energy term of the mesons is given by

$$\frac{1}{2}\partial_\mu\sigma\partial^\mu\sigma = \frac{f_\pi}{4}tr(\partial_\mu U^+ \partial^\mu U) \tag{140}$$

Next, we realize that nucleon-meson coupling term can be written as

$$\begin{aligned}
-g_\pi\left(\bar{\psi}\psi\,\sigma + \bar{\psi}\gamma_5\vec{\tau}\psi\,\vec{\pi}\right) &= -g_\pi\bar{\psi}\left[f_\pi\left(\cos(\frac{\Phi}{f_\pi}) + i\gamma_5\vec{\tau}\hat{\Phi}\sin(\frac{\Phi}{f_\pi})\right)\right]\psi \\
&= -g_\pi\bar{\psi}\left(f_\pi e^{i\gamma_5\frac{\vec{\tau}\vec{\Phi}(x)}{f_\pi}}\right)\psi \\
&= -g_\pi f_\pi \bar{\psi}\Lambda\Lambda\psi \tag{141}
\end{aligned}$$

where we have defined

$$\Lambda \equiv e^{i\gamma_5\frac{\vec{\tau}\vec{\Phi}(x)}{2f_\pi}} \tag{142}$$

If we now redefine the nucleon fields

$$\psi_W = \Lambda\psi \tag{143}$$

$$\Rightarrow \bar{\psi}_W = \psi^+\Lambda^+\gamma^0 \stackrel{\{\gamma_0,\gamma_5\}=0}{=} \psi^+\gamma^0\Lambda = \bar{\psi}\Lambda \tag{144}$$

the interaction term (141) can be simply written as

$$-g_\pi f_\pi \bar\psi \Lambda\Lambda\psi = -g_\pi f_\pi \bar\psi_W \psi_W = -M_N \bar\psi_W \psi_W \tag{145}$$

where we have used the Goldberger-Treiman relation (62). In terms of the new fields, $\psi_W$, the entire interaction term as been reduced to the nucleon mass term. If we want to identify the nucleons with the redefined fields $\psi_W$ we also have to rewrite the nucleon kinetic energy term in terms of those fields.

$$\bar\psi\, i\slashed\partial\, \psi = \bar\psi_W \Lambda^+ i\slashed\partial\, \Lambda^+ \psi_W \tag{146}$$

Since $\Lambda$ is space-dependent through the fields $\Phi(x)$, the derivative also acts on $\Lambda$, giving rise to additional terms. After some straightforward algebra, one finds

$$\bar\psi_W \Lambda^+ i\slashed\partial\, \Lambda^+ \psi_W = \bar\psi_W \left(i\slashed\partial + \gamma^\mu V_\mu + \gamma^\mu \gamma_5 A_\mu\right)\psi_W \tag{147}$$

with

$$V_\mu = \frac{1}{2}\left[\xi^+ \partial_\mu \xi + \xi \partial_\mu \xi^+\right] \tag{148}$$

$$A_\mu = \frac{i}{2}\left[\xi^+ \partial_\mu \xi - \xi \partial_\mu \xi^+\right] \tag{149}$$

$$\xi = e^{i\frac{\vec\tau \vec\Phi(x)}{2f_\pi}} \Rightarrow U = \xi\xi \tag{150}$$

We do not need to transform the potential of the linear sigma-model, $V(\pi,\sigma)$, since it vanishes on the chiral circle due to the constraint condition (135). Putting everything together, the Lagrangian of the nonlinear sigma-model, which is often referred to as the Weinberg-Lagrangian, reads in the above variables

$$\mathcal{L}_W = \bar\psi \left(i\slashed\partial + \gamma^\mu V_\mu + \gamma^\mu \gamma_5 A_\mu - M_N\right)\psi + \frac{f_\pi}{4} tr(\partial_\mu U^+ \partial^\mu U) \tag{151}$$

were we have dropped the subscript from the nucleon fields. Clearly, this Lagrangian depends nonlinearly on the fields $\vec\Phi$. It is instructive to expand the Lagrangian for small fluctuations $\Phi/f_\pi$ around the ground state. This gives

$$\begin{aligned}\mathcal{L}_W &\simeq \bar\psi(i\slashed\partial - M_N)\psi + \frac{1}{2}(\partial_\mu \vec\Phi)^2 \\ &+ \frac{1}{2f_\pi}(\bar\psi \gamma_\mu \gamma_5 \vec\tau \psi)\partial^\mu \vec\Phi - \frac{1}{4f_\pi^2}(\bar\psi \gamma_\mu \vec\tau \psi)\cdot\left(\vec\Phi \times (\partial^\mu \vec\Phi)\right)\end{aligned} \tag{152}$$

where $\vec\Phi$ is now to be identified with the pion. Comparing with the linear sigma-model, the $\sigma$-field has disappeared and the coupling between nucleons and pions has been changed to a pseudo-vector-one, involving the derivatives (momenta) of the pion-field. In addition, an explicit isovector coupling-term has emerged. From this Lagrangian it is immediately clear that the s-wave pion nucleon scattering amplitudes vanishes in the chiral limit, because all couplings involve the pion four-momentum, which at threshold is zero in case of massless pions. Thus, the important cancelation between the nucleon pole-diagrams and the $\sigma$-exchange diagram, which we found in the linear sigma-model, has been moved into the derivative coupling of the pion through the above transformations.

On the level of the expanded Lagrangian (152), the explicit breaking of chiral symmetry is introduced by an explicit pion mass term. Consequently corrections to the scattering lengths due to the nucleon pole diagrams should be of the order of $m_\pi^2$, since two derivative couplings are involved. However, the coupling $\delta \mathcal{L} = -\frac{1}{4 f_\pi^2}(\bar\psi \gamma_\mu \vec\tau \psi) \cdot \left( \vec\Phi \times (\partial^\mu \vec\Phi) \right)$, which contributes to first order to the isospin-odd amplitude, should give rise to a term $\sim \frac{m_\pi}{f_\pi^2}$ in agreement with our previous findings (122). Not too surprisingly one finds, that the above Lagrangian gives exactly the same results for the scattering-length as the linear sigma-model, except, that correction $\sim \frac{1}{m_\sigma^2}$ are absent, because we have assumed that the mass of the $\sigma$-meson is infinite. However, the full Lagrangian (151) would give rise to many more terms, if we expand to higher orders in the fields $\Phi$, which then would lead to loops etc. How to control these corrections in a systematic fashion will be the subject of the following section, where we discuss the ideas of chiral perturbation theory.

# 5  Basic ideas of Chiral Perturbation Theory

In the previous sections we were concerned with the most simple chiral Lagrangian in order to see how chiral symmetry enters into the dynamics. As we have already pointed out, many more chirally invariant terms terms can be included into the Lagrangian and thus we need some scheme which tells what to include and what not. This scheme is provided by chiral perturbation theory.

Roughly speaking, the essential idea of chiral perturbation theory is to realize that at low energies the dynamics should be controlled by the lightest particles, the pions, and the symmetries of QCD, chiral symmetry. Therefore, s-matrix elements, i.e. scattering amplitudes, should be expandable in a Taylor-series of the pion-momenta and masses, which is also consistent with chiral symmetry. This scheme will be valid until one encounters a resonance, such as the $\rho$-meson, which corresponds to a singularity of the s-matrix. Practically speaking, above the resonance, a Breit-Wigner distribution cannot be expanded in a Taylor series.

It is not too surprising that such a scheme works. Imagine, we did not know anything about QED. We still could go ahead and parameterize the, say, electron-proton scattering amplitude in powers of the momentum transfer $t$. In this case the Taylor coefficients would be related to the total charge, the charge radius etc. With this information we could write down an effective proton-electron Lagrangian, where the couplings are fixed by the above Taylor-coefficients, namely the charge and the charge- radius. This effective theory will, of course, reproduce the results of QED up to the order, which has been fixed by experiment. It is in this sense, the effective Lagrangian, obtained in chiral perturbation theory, should be understood; namely as a method of writing s-matrix elements to a given order in pion-momentum/mass. And to the order considered, the the effective Lagrangian obtained with chiral-perturbation theory should be equivalent with QCD [13, 14].

It should be stressed, that chiral perturbation theory is not a perturbation theory in the usual sense, i.e., it is not a perturbation theory in the QCD-coupling constant. In this respect, it is actually a nonperturbative method, since it takes already infinitely many order of the QCD coupling constant in order to generate a pion. Instead, as already pointed out, Chiral perturbation theory is an expansion of the s-matrix elements in terms of pion-momenta/masses.

From the above arguments one could get the impression, that chiral perturbation theory has no predictive power, since it represents simply a power expansion of measured scattering amplitudes. Although this may true in some cases, one could easily imagine that one fixes the effective Lagrangian from some experiments and then is able to calculate other observables. For example, imagine that the effective pion-nucleon interaction has been fixed from pion nucleon-scattering experiments. This interaction can then be used to calculate e.g. the photo-production of pions.

To be specific, let us discuss the case of pure pionic interaction, i.e. without any nucleons. As pointed out in the previous section, chiral invariance requires that the effective Lagrangian has to be build from structures involving $U^+U$ (138) such as

$$tr(\partial_\mu U^+ \partial^\mu U), \ tr(\partial_\mu U^+ \partial^\mu U)tr(\partial_\mu U^+ \partial^\mu U), \ tr[(\partial_\mu U^+ \partial^\mu U)^2], \ \ldots \quad (153)$$

Furthermore, each $U = e^{i\frac{\vec{\tau}\vec{\Phi}(x)}{f_\pi}}$ contains any power of the pion-field $\Phi$, which may give rise to loops etc. To specify, which of the above terms should be included into the effective Lagrangian and how much each term should be expanded in terms of the pion field, one has to count the powers of pion momenta contributing to the desired process (scattering amplitude).

Consider a given Feynman-diagram contributing to the scattering amplitude. It will have a certain number $L$ of loops, a certain number $V_i$ of vertices of type $i$ involving $d_i$ derivatives of the pion field an a certain number of internal lines $I_p$. The power $D$ of the pion momentum $q$, this diagram will have at the end, can be determined as follows:

- each loop involves an integral over the internal momenta $\int d^4q \sim q^4$

- each internal pion line corresponds to a pion propagator, and thus contributes as $\frac{1}{q^2}$

- each vertex $V_i$ involving $d_i$ derivatives of the pion field, contributes like $q^{d_i}$

Consequently, the total power of $q$, $q^D$ is given by

$$D = 4L - 2I_P + \sum_i V_i d_i \quad (154)$$

This can be simplified by using the general relation between the numbers of loops, internal lines and vertices of a given diagram

$$L = I_p - \sum_i V_i + 1 \quad (155)$$

to give

$$D = 2 + 2L + \sum_i V_i(d_i - 2) \quad (156)$$

With this formula we can determine to which order of the Taylor expansion of the scattering amplitude a given diagram contributes.

In order to see how this counting rule leads to an effective Lagrangian of a given order, we best study the simple example of pion-pion scattering. Since $U^+U = 1$ does not

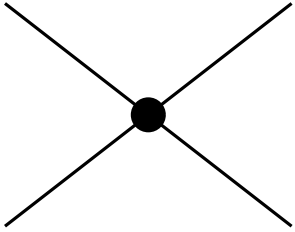

Figure 5: Leading order diagram for $\pi$-$\pi$ scattering.

contribute to the dynamics, the simplest contribution to the effective Lagrangian is given by

$$\mathcal{L}_2 = \frac{f_\pi}{4} tr(\partial_\mu U^+ \partial^\mu U) \tag{157}$$

where the subscript denotes the number of derivatives involved. Since we are discussing pion-pion scattering, we have to expand at least up to fourth order in the pion fields,

$$\mathcal{L}_2 = \frac{1}{2}(\partial_\mu \Phi)^2 + \frac{1}{6f_\pi^2}\left[(\Phi \partial_\mu \Phi)^2 - \Phi^2(\partial_\mu \Phi \partial^\mu \Phi)\right] + \mathcal{O}(\Phi^6) \tag{158}$$

where the second term contributes to the pion-pion scattering amplitude. Although this term has two contributions, for the purposes of power counting, the second term may be considered as one vertex function, because both contributions have the same number of derivatives. Thus, to lowest order, we have just one diagram, which is shown in fig. (5). It has no loops, $L = 0$, and the vertex function carries two derivatives of the pion field. Using the above counting rule (156), the order of this diagram is $D = 4$.

We can easily convince ourselves that there are no more terms contributing to this order. Including terms into the Lagrangian with four derivatives of the pions field such as e.g. $tr[(\partial_\mu U^+ \partial^\mu U)^2]$ immediately leads to $D \leq 6$. Also expanding the above Lagrangian (157) up to sixth order in the pion field leads to $D \leq 6$, because two of the pion fields have to be combined into a loop, since we are only considering a process with four external pions.

Obviously, the order of the effective Lagrangian depends on the process under consideration. Whereas a term involving six pion fields contributes to the order $D \leq 6$ to pion-pion scattering, it would contribute to order $D = 4$ to a process with three initial and three final pions. Of course, having realized, that we are actually parameterizing s-matrix elements, this is not such a surprise.

As already mentioned, to order $D = 6$ we have contributions from different sources. First of all, form higher derivative terms in the Lagrangian and secondly, from the expansion to higher order in the pion fields, giving rise to loops. The beauty of chiral perturbation theory is, that the effects of loops can be systematically be absorbed into renormalized couplings and masses. For details see e.g [15].

By now, the astute reader will have asked himself: How do I know, that a momentum is small, or in other words, what is the expansion scale? There are several answers on the market. Georgi [16] argues, based on renormalization arguments, that the scale should be $4\pi f_\pi \sim 1\,\text{GeV}$, whereas others argue [17, 15], that the mass of the lowest lying resonance should give the scale, since this is the energy, where the entire game seizes to work. This seems to be a reasonable argument and, assuming that there is no $\sigma$-meson of mass $\sim 500\,\text{MeV}$, the mass of the $\rho$-meson should provide a reasonable benchmark.

So far we have worked in the chiral limit, i.e. assuming that the pion mass vanishes. The explicit breaking of chiral symmetry is introduced by terms of the form $\sim tr(U^+ + U)$ and and the simplest symmetry breaking is

$$\delta \mathcal{L}_{X\chi SB} = \frac{f_\pi^2 m_\pi^2}{4} tr(U^+ + U) \simeq 4 - \frac{1}{2} m_\pi^2 \Phi^2 + \mathcal{O}(\Phi^4) \tag{159}$$

which to leading order in the pion-fields corresponds to a pion mass-term (the constant term does not contribute to the dynamics). Again, one can have many symmetry breaking term involving the above structure, such as

$$tr(U^+ + U), \quad tr(\partial_\mu U^+ \partial^\mu U) tr(U^+ + U) \ldots \tag{160}$$

so that an ordering scheme is necessary. Therefore, in the realistic case of explicit chiral symmetry breaking, the scattering amplitudes are not only expanded in terms of the pion momenta but also in terms of the pion masses. The counting-rule is the same as given above (156), where $d_i$ now gives the number of derivatives *and* pion masses of a given vertex of type $i$. The total effective Lagrangian for pion-pion scattering to order $D = 4$ is then given by

$$\mathcal{L}_2^{(4)} = \frac{1}{2}(\partial_\mu \Phi)^2 + \frac{1}{2} m_\pi^2 \Phi^2 \frac{1}{6 f_\pi^2} \left[ (\vec{\Phi} \cdot \partial_\mu \vec{\Phi})^2 - \Phi^2 (\partial_\mu \vec{\Phi} \cdot \partial^\mu \vec{\Phi}) \right] + \frac{m_\pi^2}{24 f_\pi^2} (\vec{\Phi} \cdot \vec{\Phi})^2 \tag{161}$$

In principle the 'adjustable' parameters of this Lagrangian are the pion-mass and the pion-decay constant, which have to be fixed to the experimental values.

The resulting pion-pion scattering length and volumes are then given by [18]

$$a_0^0 = \frac{7 m_\pi}{32 \pi f_\pi^2}, \quad a_0^2 = -\frac{m_\pi}{16 \pi f_\pi^2}, \quad a_1^1 = \frac{1}{24 \pi f_\pi^2 m_\pi} \tag{162}$$

where the subscript denotes the angular momentum and the superscript the isospin of the amplitude. As shown in table (1) [15], the leading order results agree reasonably well with experiment and are improved by the next to leading order corrections. Apparently we do not find perfect agreement with experiment even for the s-wave scattering lengths, although already to leading order we haven taken into account terms quadratic in the momenta, so that higher orders in the pion momentum will not improve the situation. However, remember, that we not only expand in terms of the pion momenta, but also, as a result of the explicit symmetry breaking, in terms of the pion mass, which in principle can contribute to any order to the s-wave scattering length.

|   | Experiment | Lowest Order | First Two Orders |
|---|---|---|---|
| $a_0^0 m_\pi$ | $0.26 \pm 0.05$ | $0.16$ | $0.20$ |
| $a_0^2 m_\pi$ | $-0.028 \pm 0.012$ | $-0.045$ | $-0.041$ |
| $a_1^1 m_\pi^3$ | $0.038 \pm 0.002$ | $0.030$ | $0.036$ |

Table 1: Pion-pion scattering length

As already pointed out in the beginning of this section, chiral perturbation theory, or more precisely, the expansion in momenta breaks down, once we get close to a resonance. This one easily understands by looking at the Breit-Wigner formula for the scattering amplitude involving a resonance.

$$f(E) \sim \frac{\Gamma/2}{E_R - E - i\Gamma/2} \quad (163)$$

For energies, which are small compared to the resonance energy, $E \ll E_R$ this amplitude may be expanded in terms of a power series and the concept of chiral perturbation theory works well

$$f(E) \sim \frac{\Gamma/2}{E_R}\left(1 + \frac{E + i\Gamma/2}{E_R} + \ldots\right); \quad E \ll E_R \quad (164)$$

However, once we get close to the resonance-energy, we need to expand to higher and higher order until at $E \geq E_R$ the power-series in $E$ seizes to converge. To be specific, we expect that in the the isovector p-wave channel, which is dominated by the $\rho$-meson resonance, the chiral perturbation expansion should fail for energies $E \sim m_\rho$.

Finally, let us include the nucleons into the chiral counting. Naively, one would think, that this should destroy the entire concept, because the nucleon has a large mass, which is of the order of the expansion scale. However, since at low energies the scattering amplitude may also be calculated in a nonrelativistic framework, we do not expect the nucleon mass to enter directly, but, to leading order, only via the kinetic energy $\sim \frac{p^2}{2M_N}$, which is small compared to that of the pion at the same momentum. Therefore, chiral perturbation theory should also work with nucleons present (for details see. [19]). The above argument can be formalized by realizing that the nucleon only enters the amplitudes through the nucleon propagator (see e.g. the results of section (3.3)). At low momenta, the nucleon propagator contributing to diagram (a) of fig. (4) can be written as

$$\frac{\gamma_\mu(p^\mu + q^\mu) + M_N}{(p+q)^2 - M_N^2} \simeq \frac{\gamma_0 M_N + M_N}{2 M_N q} = \frac{\Lambda}{q}(1 + \mathcal{O}(\frac{q}{M_N})) \qquad (165)$$

where

$$\Lambda = \frac{\gamma_0 M_N + M_N}{2 M_N} = \begin{pmatrix} 1 & 0 \\ 0 & 0 \end{pmatrix} \qquad (166)$$

projects on positive energy states. Hence, to leading order, each nucleon propagator contributes like $\frac{1}{q}$ to the power of pion momentum of the scattering amplitude. This leads to the following counting rule, which now also includes the nucleons [19]

$$D = 2 + 2L - \frac{1}{2} E_N + \sum_i V_i (d_i + \frac{1}{2} n_i - 2) \qquad (167)$$

Here the notation is as in equ. (156) and $E_N$ denotes the number of external nucleon lines and $n_i$ the number of nucleon fields of vertex $i$, which is typically $n_i = 2$.

For the simple nucleon-pole diagram using pseudovector coupling we thus would have: $L = 0$, $E_N = 2$, $d = 1$, $n = 2$ such that, $d + \frac{1}{2} n - 2 = 0$ and, $D = 1$.

On top of the expansion in terms of pion-momenta and pion masses, from equ. (165) we, therefore, also have an expansion in the velocity of the nucleons $v \sim \frac{q}{M_N}$. This is carried out in a systematic fashion in the so called Heavy-Baryon Chiral-Perturbation Theory, as introduced by Jenkins and Manohar [20]. This approach essentially corresponds to a systematic nonrelativistic expansion for the nucleon wave-function, on the basis that the nucleon (baryon) is heavy compared to the momenta involved. We should mention, that the effect of the nucleon can also be included in a fully covariant fashion as discussed by Gasser et al. [21].

Including the nucleon gives rise to additional structures which explicitly break the chiral symmetry, such as

$$\delta \mathcal{L} = a \, tr(U^+ + U) \bar{\psi} \psi \simeq a(1 - \frac{\phi^2}{2 f_\pi^2}) \bar{\psi} \psi \qquad (168)$$

To leading order, this is just a contribution to the nucleon mass, which allows us to identify the coefficient $a$ with the sigma-term $\Sigma_{\pi N}$ (100)

$$\delta \mathcal{L} = -\Sigma_{\pi N} \, tr(U^+ + U) \bar{\psi} \psi \simeq -\Sigma_{\pi N} \bar{\psi} \psi + \frac{\Sigma_{\pi N}}{2 f_\pi^2} \bar{\psi} \psi \, \phi^2 \qquad (169)$$

The next to leading term in the above expression is an *attractive* interaction between pion and nucleon, which contributes to the order $D = 2$ to the amplitude. This term by itself is quite large and would lead to a wrong prediction for the s-wave pion-nucleon amplitude. However, there are additional terms contributing to the same order, which in the heavy-fermion expansion comes from the nucleon-pole diagrams. The coefficients of these terms then need to be chosen such, that the resulting scattering length acquire the small value observed in experiment [22].

# 6 Applications

In this last section, we want to discuss a few applications of chiral symmetry relevant for the physics of dense and hot matter. First we briefly address the issue of in medium masses of pions and kaons. Then we will discuss the temperature and density dependence of the quark condensate. We will conclude with some general remarks on the properties of vector mesons in matter.

## 6.1 Pion and kaon masses in dense matter

Changes of the pion mass in the nuclear medium should show up in the iso-scalar pion s-wave optical potential. To leading order in the density this is related to the s-wave iso-scalar scattering-length $a_0^+$ by [11]

$$2\omega U = -4\pi (1 + \frac{m_\pi}{M_N}) a_0^+ \rho \qquad (170)$$

where $\omega$ is the pion energy. Since the s-wave iso-scalar scattering length is small, as a result of chiral symmetry, and slightly repulsive, we predict a small increase of the pion mass in the nuclear medium, which at nuclear matter amounts to $\Delta m_\pi \simeq 5 MeV$. One arrives at the same result by evaluating the effective Lagrangian, as obtained from chiral perturbation theory, at finite density [23, 22]. This is not surprising since the s-wave iso-scalar amplitude is used to fix the relevant couplings.

In case of the kaons, which can also be understood as Goldstone bosons of an extended $SU(3) \times SU(3)$ chiral symmetry, some interesting features occur. Chiral perturbation theory predicts a repulsive s-wave scattering length for $K^+$-nucleon scattering and a large attractive one for $K^-$ [24, 23]. Using the above relation for the optical potential (170) this led to speculations about a possible s-wave kaon condensate in dense matter [24, 25] with rather interesting implications for the structure and stability of neutron stars [26, 27]. Experimentally, however, one finds that the iso-scalar s-wave scattering length for the $K^-$ is repulsive, calling into question the results from chiral perturbation theory. The resolution to this puzzle is the presence of the $\Lambda(1405)$ resonance just below the kaon-nucleon threshold. This resonance, which has not been taken into account in the chiral perturbation analysis, gives a large repulsive contribution to the scattering amplitude at threshold. Does that mean, that chiral perturbation theory failed? Yes and no. Yes, because, as already pointed out, it is not able to generate any resonances and thus leads to bad predictions in the neighborhood of the resonance[14]. No, because it predicts a strong attraction between the proton and the $K^-$, which, if iterated to infinite order can generate the $\Lambda(1405)$-resonance as a bound state in the continuum [29]( in the continuum, because the $\Lambda(1405)$ decays into $\Sigma\pi$).

---
[14]Lee et al. [28] have attempted to include the $\Lambda(1405)$ as an explicit state in a chiral perturbation theory analyses of the kaon-nucleon scattering length. While this approach may be a reasonable thing to do phenomenologically, it appears to be beyond the original philosophy of chiral perturbation theory.

This situation is well known from nuclear physics. The proton-neutron scattering length in the deuteron channel is repulsive although the proton-neutron interaction is attractive. The reason is, that in this channel a bound state can be formed, the deuteron, which gives rise to a strong repulsive contribution to the scattering amplitude at threshold.

To carry this analogy further, we know that in nuclear matter the deuteron has disappeared, essentially due to Pauli-blocking, revealing the true, attractive nature of the nuclear interaction. As a result we have an attractive mean field potential for the nucleons. Similarly, one can argue [30], that the $\Lambda(1405)$, if it is a $K^-$-proton bound state, should eventually disappear, resulting in an attractive s-wave optical potential for the $K^-$ in nuclear matter. Indeed, an analysis of $K^-$ atoms [31], shows, that the optical potential turns attractive already at rather low densities $\rho \leq 0.5\rho_0$. Extrapolated to nuclear matter density the extracted optical potential would be as deep as $-200\,\mathrm{MeV}$, in reasonable agreement with the predictions from chiral perturbation theory.

## 6.2 Change of the quark-condensate in hot and dense matter

### 6.2.1 Temperature dependence

One of the applications of chiral perturbation theory relevant to the physics of hot and dense matter, is the calculation of the temperature dependence of the quark condensate. Here we just want derive the leading order result. A detailed discussion, which includes also higher order corrections can be found in ref. [32]. The basic idea, is to realize that the operator of the quark-condensate, $\bar{q}q$, enters into the QCD-Lagrangian via the quark mass term. Thus, we may write the QCD-Hamiltonian as

$$H = H_0 + m_q \bar{q}q \tag{171}$$

The quark condensate at finite temperature is then given by the following statistical sum

$$<\bar{q}q>_T = \frac{\sum_i <i|\bar{q}q\, e^{-H/t}|i>}{\sum_i <i|e^{-H/T}|i>} \tag{172}$$

Since $\partial H/\partial m_q = \bar{q}q$ this can be written as

$$<\bar{q}q>_T = T\frac{\partial}{\partial m_q}\ln Z(m_q) \tag{173}$$

where the partition function $Z$ is given by $Z = \sum_i <i|e^{-H/T}|i>$.

In chiral perturbation theory we do not calculate the partition function of QCD, but rather that of the effective Lagrangian. To make contact with the above relations, we utilize the Gell-Mann Oakes Renner relation (103). To leading order in the pion mass the derivative with respect to the quark mass, therefore, can be written as

$$\frac{\partial}{\partial m_q} = -\frac{<0|\bar{q}q|0>}{f_\pi^2} \frac{\partial}{\partial m_\pi^2} \tag{174}$$

Next to leading order contributions arise, among others, from the quark-mass dependence of the vacuum condensate.

To leading order the partition function is simply given by that of a noninteracting pion gas

$$\ln Z = \ln Z_0 + \ln Z_{\pi-\text{gas}} = \ln Z_0 + \frac{3}{(2\pi)^3} \int d^3p \, \ln(1 - \exp(-E/T)) \tag{175}$$

where $Z_0$ stands for the vacuum contribution, which we, of course, cannot calculate in chiral perturbation theory, since we are only concerned about fluctuation around that vacuum. Thus the temperature dependence of the quark condensate in the chiral limit is given by

$$\begin{aligned}
<\bar{q}q>_T &= <0|\bar{q}q|0> - \frac{<0|\bar{q}q|0>}{f_\pi^2} \frac{\partial}{\partial m_\pi^2} Z_{\pi-\text{gas}} \bigg|_{m_\pi \to 0} \\
&= <0|\bar{q}q|0> (1 - \frac{T^2}{8f_\pi^2})
\end{aligned} \tag{176}$$

Thus to leading order, the quark condensate drops like $\sim T^2$, i.e. at low temperatures the change in the condensate is small.

Corrections include the effect of pion interactions, which in the chiral limit are proportional to the pion momentum and thus contribute to higher orders in the temperature. Including contributions up to three loops, one finds see e.g. [32]

$$\frac{<\bar{q}q>_T}{<\bar{q}q>_0} = 1 - c_1 \left(\frac{T^2}{8f_\pi^2}\right) - c_2 \left(\frac{T^2}{8f_\pi^2}\right)^2 - c_3 \left(\frac{T^2}{8f_\pi^2}\right)^3 \ln(\frac{\Lambda_q}{T}) + \mathcal{O}(T^8) \tag{177}$$

For $N_f$ flavors of massless quarks the coefficients are given in the chiral limit by

$$c_1 = \frac{2}{3} \frac{N_f^2 - 1}{N_f} \qquad c_2 = \frac{2}{9} \frac{N_f^2 - 1}{N_f^2} \qquad c_3 = \frac{8}{27} (N_f^2 + 1) N_f \tag{178}$$

The scale $\Lambda_q$ can be fixed from pion scattering data to be $\Lambda_q = 470 \pm 110 \, \text{MeV}$. In fig. (6) we show the temperature dependence of the quark-condensate as predicted by the above formula. Currently, lattice gauge calculations predict a critical temperature $T_c \simeq 150 \, \mathrm{MeV}$, above which the quark condensate has disappeared. At this temperature chiral perturbation theory predicts only a drop of about 50 %, which gets even smaller once pion masses are included [32]. However, we do not expect chiral perturbation to work well close to the critical temperature. The strength of this approach is at low temperatures. The prediction, that to leading order the condensate drops quadratic in the temperature is a direct consequence of chiral symmetry and can be used to check chiral models as well as any other conjectures involving the change of the quark-condensate, such as e.g. the change of hadron masses.

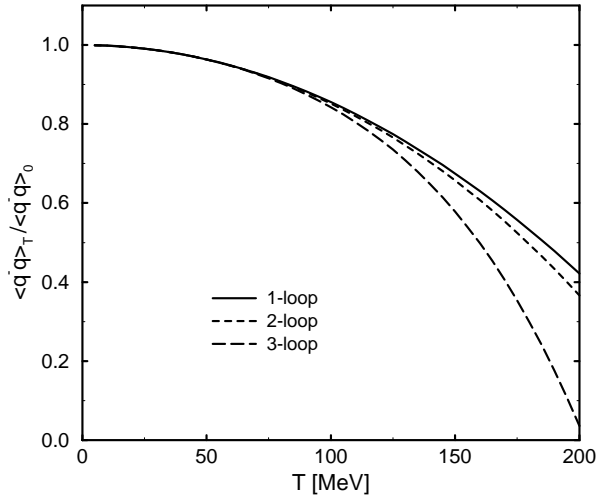

Figure 6: Temperature dependence of the quark condensate from chiral perturbation theory (chiral limit).

### 6.2.2 Density dependence

For low densities, the density dependence of the quark condensate can also be determined in a model independent way[15]. We expect that to leading order in density the change in the quark condensate is simply given by the amount of quark condensate in a nucleon multiplied by the nuclear density,

$$<\bar{q}q>_\rho = <\bar{q}q>_0 + <N|\bar{q}q|N> \rho + \text{higher orders in } \rho \qquad (179)$$

---

[15] Again, we give a heuristic argument. A rigorous derivation based on the Hellmann-Feynman theorem can be found e.g. in [33, 34].

All we need to know is the matrix element of $\bar{q}q$ between nucleon states. This matrix element, however, enters into the pion-nucleon sigma-term (104)

$$< N|\bar{q}q|N > = \frac{\Sigma_{\pi N}}{m_q} = \Sigma_{\pi N} \frac{<\bar{q}q>_0}{m_\pi^2 f_\pi^2} \qquad (180)$$

where we also have made use of the GOR-relation (103), namely $m_q = \frac{m_\pi^2 f_\pi^2}{<\bar{q}q>_0}$. Thus we predict, that the quark condensate drops *linearly* with density, as compared to the quadratic temperature dependence found above

$$<\bar{q}q>_\rho = <\bar{q}q>_0 \left(1 - \frac{\Sigma_{\pi N}}{m_\pi^2 f_\pi^2}\rho + \ldots\right) \qquad (181)$$

Corrections to higher order in density arrise, among others, from nuclear binding effects. These have been estimated by Brockmann [35] to be at most of the order of 15 % for denities up to twice nuclear-matter density. Assuming a value for the sigma term of $\Sigma_{\pi N} \simeq 45$ MeV we find that the condensate has dropped by about 35 % at nuclear matter density

$$<\bar{q}q>_\rho = <\bar{q}q>_0 \left(1 - 0.35\frac{\rho}{\rho_0}\right) \qquad (182)$$

Thus, finite density is very efficient in reducing the quark condensate and we should expect that any in medium modification due to a dropping quark condensate should already be observable at nuclear matter density. The above findings also suggest, that chiral restoration, i.e. the vanishing of the quark-condensate, is best achieved in heavy ion collisions at bombarding energies, which still lead to full stopping of the nuclei.

## 6.3 Masses of vector mesons

Finally, let us briefly discuss what chiral symmetry tells us about the masses of vector mesons in the medium. Vector mesons, such as the $\rho$-meson, are of particular interest, because they decay into dileptons. Therefore, possible changes of their masses in medium are accessible to experiment.

Using current algebra and PCAC, Dey et al. [36] could show, that at finite temperature the mass of the rho-meson does not change to order $T^2$. Instead to order $T^2$ the vector-correlation function gets an admixture from the axial-vector correlation function

$$C_V(T) = (1 - \epsilon)C_V(T = 0) + \epsilon C_A(T = 0) \qquad (183)$$

with $\epsilon = \frac{T^2}{6f_\pi^2}$. The imaginary part of this vector-correlation function is directly related to the dilepton-production cross-section. As depicted in fig. (7), the above result, therefore, predicts that to leading order in the temperature, the dilepton invariant mass spectrum develops a peak at the mass of the $a_1$-meson in addition to that at the mass of the $\rho$. At the same time, the contribution at the $\rho$-peak is reduced in comparison to the free case. Furthermore, the position of the peaks is not changed to this order in temperature. This general result is also confirmed by calculations in chiral models, which have been extended to include vector mesons [37, 38]. Notice, that the above finding also rules out that the mass of the $\rho$-meson scales linearly with the quark condensate, because previously (see section 6.2.1) we found that the quark condensate already drops to order $T^2$, whereas the mass of the $\rho$ does not change to this order.

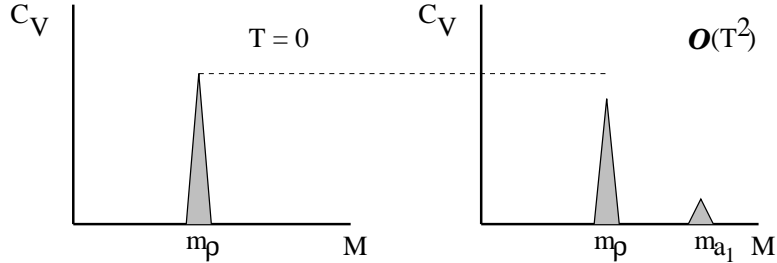

Figure 7: Vector-spectral functions at $T = 0$ and to leading order in temperature as given by equ. (183).

Corrections to higher order in the temperature, however, are not controlled by chiral symmetry alone and, therefore, one finds model dependencies. Pisarski [38] for instance predicts in the framework of a linear sigma model with vector mesons, that to order $T^4$ the mass of the $\rho$ decreases and that of the $a_1$ increases. Song [37], on the other, uses a nonlinear $\sigma$-model and finds the opposite, namely, that the $\rho$ goes up and the $a_1$ goes down. At the critical temperature, both again agree qualitatively in that the masses of $a_1$ and $\rho$ become degenerate at a value which is roughly given by the average of the vacuum masses $\simeq 1\,GeV$. This agreement, again, is a result of chiral symmetry.

At and above the critical temperature, where chiral symmetry is not anymore spontaneously broken, chiral symmetry demands that the vector and axial vector correlation functions are the same. One way to realize that is by the having the same masses for the vector ($\rho$) and axial-vector ($a_1$). However, this is not the only possibility! As nicely discussed in a paper by Kapusta and Shuryak [39], there are at least three qualitatively different possibilities, which are sketched in fig. (8).

1. The masses of $\rho$ and $a_1$ are the same. In this case, clearly the vector and axial vector correlation functions are the same. Note, however, that we cannot make any statement about the value of the common mass. It may be zero, as suggested by some people, it may be somewhere in between the vacuum masses, as the chiral

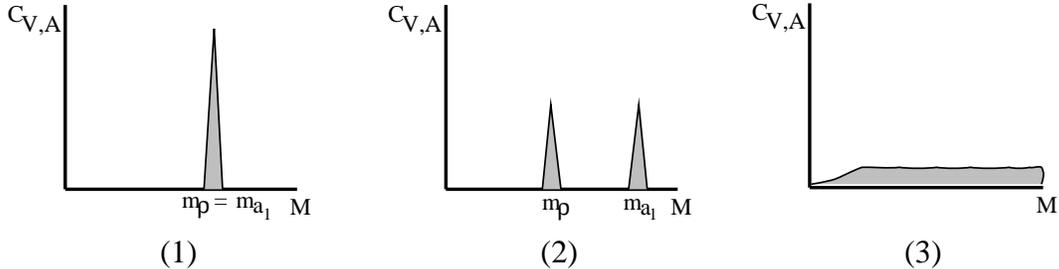

Figure 8: Several possibilities for the vector and axial-vector spectral functions in the chirally restored phase.

    models seem to predict and it my be even much larger than the mass of the $a_1$.

2. We may have a complete mixing of the spectral functions. Thus, both the vector and axial-vector spectral functions have peaks of equal strength at both the mass of the $\rho$ and the mass of the $a_1$, leading to two peaks of equal strength in the dilepton spectrum (modulo Boltzmann-factors of course). One example would be given by the low temperature result (183) with the mixing parameter $\epsilon(T_c) = \frac{1}{2}$. Using the low temperature result for $\epsilon = \frac{T^2}{6f_\pi^2}$ would give a critical temperature of $T_c = \sqrt{3} f_\pi \simeq 164\, MeV$, which is surprisingly close to the value given by recent lattice calculations.

3. Both spectral functions could be smeared over the entire mass range. Due to thermal broadening of the mesons and the onset of deconfinement, the structure of the spectral function may be washed out and it becomes meaningless to talk about mesonic states.

    To summarize, the only unique prediction derived from chiral symmetry (current algebra) about the temperature dependence of the $\rho$-mass, is that it does not change to order $T^2$, i.e. at low temperatures. Furthermore, at and above the critical temperature, chiral symmetry requires that the vector and axial-vector spectral functions are identical, which, however, does not necessarily imply, that both exhibit just one peak, located at the same position. Corrections of the order $T^4$ cannot be obtained from chiral symmetry alone.

    Finally let us point out, that the above findings do not rule out scenarios, which relate the mass of the $\rho$ with the temperature dependence of the bag-constant or gluon condensate, such as proposed by Pisarski [40] and Brown and Rho [41]. This ideas, however, involve concepts which go beyond chiral symmetry, such as the melting of the gluon condensate. Consequently in these scenarios, a certain behavior of the mass of the $\rho$-meson can only indirectly be brought in connection with chiral restoration.

**Acknowledgments:** I would like to thank C. Song for useful discussions concerning the effects of chiral symmetry on the vector mesons. This work was supported by the Director, Office of Energy Research, Office of High Energy and Nuclear Physics Division of the Department of Energy, under Contract No. DE-AC03-76SF00098.

# 7 Appendix: Useful references

This is a selection of references, which the author found useful in preparing these lectures. It is by no means a complete representation of the available literature.

1. D.K. Campbell, 'Chiral symmetry, pions and nuclei',
   in 'Nuclear Physics with Heavy Ions and Mesons', Volume 2, Editors: Balian, Rho and Ripka, (Les Houches XXX, 1977),North Holland.
   Comment: Very nice introduction to chiral symmetry.

2. H. Georgi, 'Weak Interactions and modern Particle Physics',
   Benjamin / Cummings, 1984
   Comment: Good introduction to the idea of chiral pert. theory, sometimes a little brief.

3. J.J. Sakurai, 'Currents and Fields',
   Chicago Univ. Press, 1969
   Comment: Nice little book about current algebra etc. These are lecture notes and, therefore, rather explicit.

4. De Alfaro, Fubini, Furlan and Rosseti, 'Currents in hadron Physics',
   North Holland, 1973.
   Comment: The current algebra 'bible'. Contains everything up to 1973 (no chiral pert. theory).

5. U. Meissner, 'Recent developments in chiral perturbation theory',
   Rep. Prog. Phys. 56 (1993) 903
   Comment: Good review about the technical aspects of chiral pert. theory.

6. H. Leutwyler, 'Principles of Chiral Perturbation Theory'
   Lectures from 'Hadrons 94" workshop, Gramado, RS, Brasil, hep-ph/9406283
   Comment: Very nice review about the conceptual aspects of chiral perturbation theory. Somewhat complementary to that of Meissner.